\documentclass[twoside]{report}
\usepackage[utf8]{inputenc}
\usepackage[english]{babel}

\usepackage{graphicx}
\graphicspath{{images/}{../images/}}

\usepackage{subfiles}

\usepackage[left=0.5in, right=0.5in, top=0.75in, bottom=0.75in]{geometry}

\usepackage{afterpage} % Blank page without messing up page numbers
\usepackage{amsthm} % Math
\usepackage{amsmath} % Math
\usepackage{amssymb} % Math
\usepackage[tikz]{bclogo} % Fancy boxes
\usepackage{booktabs} % Better tables
\usepackage[authordate, maxcitenames=1]{biblatex-chicago} % chicago bibliography style
\usepackage{caption} % Captions
\usepackage{csquotes} % For chicago bibliography style
\usepackage{csvsimple} % CSV imports
\usepackage[type={CC}, modifier={by-sa}, version={3.0}]{doclicense} % CC License
\usepackage{enumerate} % Special enumeration
\usepackage{fancyhdr} % Fancy heders
\usepackage{float} % Figure position
\usepackage[hang, bottom]{footmisc} % Advanced footer formatting
\usepackage[toc]{glossaries} % Glossary
\usepackage{graphicx} % Graphics
\usepackage[hidelinks]{hyperref} % Hyperlinksxx
\usepackage{longtable} % Long tables
\usepackage{parskip} % Custom skips
\usepackage{relsize} % Relative sizing
\usepackage{rotating} % Rotated images
\usepackage{supertabular} % Multicolumn long tables
\usepackage{wrapfig} % Text wrapping around figures
\allowdisplaybreaks[1]
\let\origdoublepage\cleardoublepage
\newcommand{\clearemptydoublepage}{%
  \clearpage
  {\pagestyle{empty}\origdoublepage}%
}

\newcommand{\citeFormat}[1]{\footnote{#1}}
\renewbibmacro*{cite:ibid}{\printtext[bibhyperref]{\bibstring[\mkibid]{ibidem}}}

\newcommand{\wrapfigadjustment}{-1.5em}

\DeclareSourcemap{ % Used when .bib/Bibliography is compiled, not when document is
    \maps{
        \map{ % Replaces '{\_}', '{_}' or '\_' with just '_'
            \step[fieldsource=url,
                  match=\regexp{\{\\\_\}|\{\_\}|\\\_},
                  replace=\regexp{\_}]
        }
        \map{ % Replaces '{'$\sim$'}', '$\sim$' or '{~}' with just '~'
            \step[fieldsource=url,
                  match=\regexp{\{\$\\sim\$\}|\{\~\}|\$\\sim\$},
                  replace=\regexp{\~}]
        }
    }
}

\fancypagestyle{plain}{
	\fancyhead{}
	\fancyfoot{}
    \fancyhead[LE,RO]{\nouppercase{\leftmark}}
    \fancyhead[C]{Learned Sectors}
    \fancyhead[RE,LO]{Weerawarana, Zhu, He}

	\fancyfoot[RE,LO]{SIT}
	\fancyfoot[C]{FE 800 - Special Research Problems}
	\fancyfoot[LE,RO]{\thepage}
}

\usepackage{multicol}
\newcommand{\innerCover}{
    \vspace*{\fill}
        \begin{center}
        \begin{minipage}{.75\linewidth}

        \doclicenseThis
        
        \vspace{2em}
        \hrule
        \vspace{2em}
        
        The underlying source code is licensed under the MIT License: \\
        \\
        \texttt{Copyright (c) 2019 Rukmal Weerawarana, Yiyi Zhu, Yuzhen He \\
        \\
        Permission is hereby granted, free of charge, to any person obtaining a copy
        of this software and associated documentation files (the "Software"), to deal
        in the Software without restriction, including without limitation the rights
        to use, copy, modify, merge, publish, distribute, sublicense, and/or sell
        copies of the Software, and to permit persons to whom the Software is
        furnished to do so, subject to the following conditions: \\
        \\
        The above copyright notice and this permission notice shall be included in all
        copies or substantial portions of the Software. \\
        \\
        THE SOFTWARE IS PROVIDED "AS IS", WITHOUT WARRANTY OF ANY KIND, EXPRESS OR
        IMPLIED, INCLUDING BUT NOT LIMITED TO THE WARRANTIES OF MERCHANTABILITY,
        FITNESS FOR A PARTICULAR PURPOSE AND NONINFRINGEMENT. IN NO EVENT SHALL THE
        AUTHORS OR COPYRIGHT HOLDERS BE LIABLE FOR ANY CLAIM, DAMAGES OR OTHER
        LIABILITY, WHETHER IN AN ACTION OF CONTRACT, TORT OR OTHERWISE, ARISING FROM,
        OUT OF OR IN CONNECTION WITH THE SOFTWARE OR THE USE OR OTHER DEALINGS IN THE
        SOFTWARE.}
        \end{minipage}
        \end{center}
    \vspace*{\fill}
    \thispagestyle{empty}
    \addtocounter{page}{-1}
    \newpage
}

\makeatletter
\def\LT@start{%
  \let\LT@start\endgraf
  \endgraf\penalty\z@\vskip\LTpre
  \dimen@\pagetotal
  \advance\dimen@ \ht\ifvoid\LT@firsthead\LT@head\else\LT@firsthead\fi
  \advance\dimen@ \dp\ifvoid\LT@firsthead\LT@head\else\LT@firsthead\fi
  \advance\dimen@ \ht\LT@foot
  \dimen@ii\vfuzz
  \vfuzz\maxdimen
    \setbox\tw@\copy\z@
    \setbox\tw@\vsplit\tw@ to \ht\@arstrutbox
    \setbox\tw@\vbox{\unvbox\tw@}%
  \vfuzz\dimen@ii
  \advance\dimen@ \ht
        \ifdim\ht\@arstrutbox>\ht\tw@\@arstrutbox\else\tw@\fi
  \advance\dimen@\dp
        \ifdim\dp\@arstrutbox>\dp\tw@\@arstrutbox\else\tw@\fi
  \advance\dimen@ -\pagegoal
  \ifdim \dimen@>\z@\vfil\break\fi
      \global\@colroom\@colht
  \ifvoid\LT@foot\else
    \global\advance\vsize-\ht\LT@foot
    \global\advance\@colroom-\ht\LT@foot
    \dimen@\pagegoal\advance\dimen@-\ht\LT@foot\pagegoal\dimen@
    \maxdepth\z@
  \fi
  \ifvoid\LT@firsthead\copy\LT@head\else\box\LT@firsthead\fi\nobreak
  \output{\LT@output}}

\def\endlongtable{%
  \crcr
  \noalign{%
    \let\LT@entry\LT@entry@chop
    \xdef\LT@save@row{\LT@save@row}}%
  \LT@echunk
  \LT@start
  \unvbox\z@
  \LT@get@widths
  \if@filesw
    {\let\LT@entry\LT@entry@write\immediate\write\@auxout{%
      \gdef\expandafter\noexpand
        \csname LT@\romannumeral\c@LT@tables\endcsname
          {\LT@save@row}}}%
  \fi
  \ifx\LT@save@row\LT@@save@row
  \else
    \LT@warn{Column \@width s have changed\MessageBreak
             in table \thetable}%
    \LT@final@warn
  \fi
  \endgraf\penalty -\LT@end@pen
  \ifvoid\LT@foot\else
    \global\advance\vsize\ht\LT@foot
    \global\advance\@colroom\ht\LT@foot
    \dimen@\pagegoal\advance\dimen@\ht\LT@foot\pagegoal\dimen@
  \fi
  \endgroup
  \global\@mparbottom\z@
  \endgraf\penalty\z@\addvspace\LTpost
  \ifvoid\footins\else\insert\footins{}\fi}

\def\LT@output{%
  \ifnum\outputpenalty <-\@Mi
    \ifnum\outputpenalty > -\LT@end@pen
      \LT@err{floats and marginpars not allowed in a longtable}\@ehc
    \else
      \setbox\z@\vbox{\unvbox\@cclv}%
      \ifdim \ht\LT@lastfoot>\ht\LT@foot
        \dimen@\pagegoal
  \advance\dimen@\ht\LT@foot
        \advance\dimen@-\ht\LT@lastfoot
        \ifdim\dimen@<\ht\z@
          \setbox\@cclv\vbox{\unvbox\z@\copy\LT@foot\vss}%
          \@makecol
          \@outputpage
      \global\vsize\@colroom
          \setbox\z@\vbox{\box\LT@head}%
        \fi
      \fi
      \unvbox\z@\ifvoid\LT@lastfoot\copy\LT@foot\else\box\LT@lastfoot\fi
    \fi
  \else
    \setbox\@cclv\vbox{\unvbox\@cclv\copy\LT@foot\vss}%
    \@makecol
    \@outputpage
      \global\vsize\@colroom
    \copy\LT@head\nobreak
  \fi}
\makeatother
\makeglossaries

\newglossaryentry{etf}
{
    name=ETF,
    description={Exchange traded fund}
}

\newglossaryentry{sector_universe}
{
    name=Sector Universe,
    description={A specific sector classification taxonomy, such as the GICS (Global Inudstry Classifciation Standard), or the ICB (Industry Classification Benchmark)}
}

\newglossaryentry{setf}
{
    name=SETF,
    description={Synthetic Exchange Traded Fund; a hypothetical asset, used in backtesting simulations by the \textit{reIndexer} research tool}
}

\newglossaryentry{reindexer}
{
    name=reIndexer,
    description={Research tool for backtest-driven evaluation of different sectorization universes, using a system of synthetic ETFs, and efficient portfolios of those synthetic ETFs}
}

\newglossaryentry{benchmark_universe}
{
    name=Benchmark Universe,
    description={For this project, the benchmark sector classification unvierse is the \textit{GICS S\&P 500 Classification}}
}

\newglossaryentry{setf_restructuring}
{
    name=SETF Restructuring Turnover,
    description={The dollar-value change of component asset turnover incurred when a synthetic ETF is restructured}
}

\newglossaryentry{portfolio_rebalancing}
{
    name=Portfolio Rebalancing Turnover,
    description={The dollar-value change of constituent SETF turnover incurred when a portfolio of SETFs is rebalanced}
}

\newglossaryentry{HCA}
{
    name=HCA,
    description={Hierarchical clustering analysis model}
}

\newglossaryentry{SIC}
{
    name=SIC,
    description={Standard Industrial Classification}
}

\newglossaryentry{NAICS}
{
    name=NAICS,
    description={North American Industry Classification System}
}

\newglossaryentry{GICS}
{
    name=GICS,
    description={Global Industry Classification System; informs the S\&P 500 Sector Classifications}
}

\newglossaryentry{WRDS}
{
    name=WRDS,
    description={Wharton Research Data Services}
}

\begin{document}

% Title page
\afterpage{\innerCover} % Blank page without messing up page numbers
\documentclass[../main.tex]{subfiles}

\begin{titlepage}

%%%%%%%%%%%%%%%%%%%%%%%%%%%%%%%%%%%%
% Modified version of template from:
% http://www.latextemplates.com/template/university-assignment-title-page
%%%%%%%%%%%%%%%%%%%%%%%%%%%%%%%%%%%%

\newcommand{\HRule}{\rule{\linewidth}{0.5mm}} % Defines a new command for the horizontal lines, change thickness here

\begin{center}

%----------------------------------------------------------------------------------------
%	HEADING SECTIONS
%----------------------------------------------------------------------------------------

\includegraphics[height=4cm]{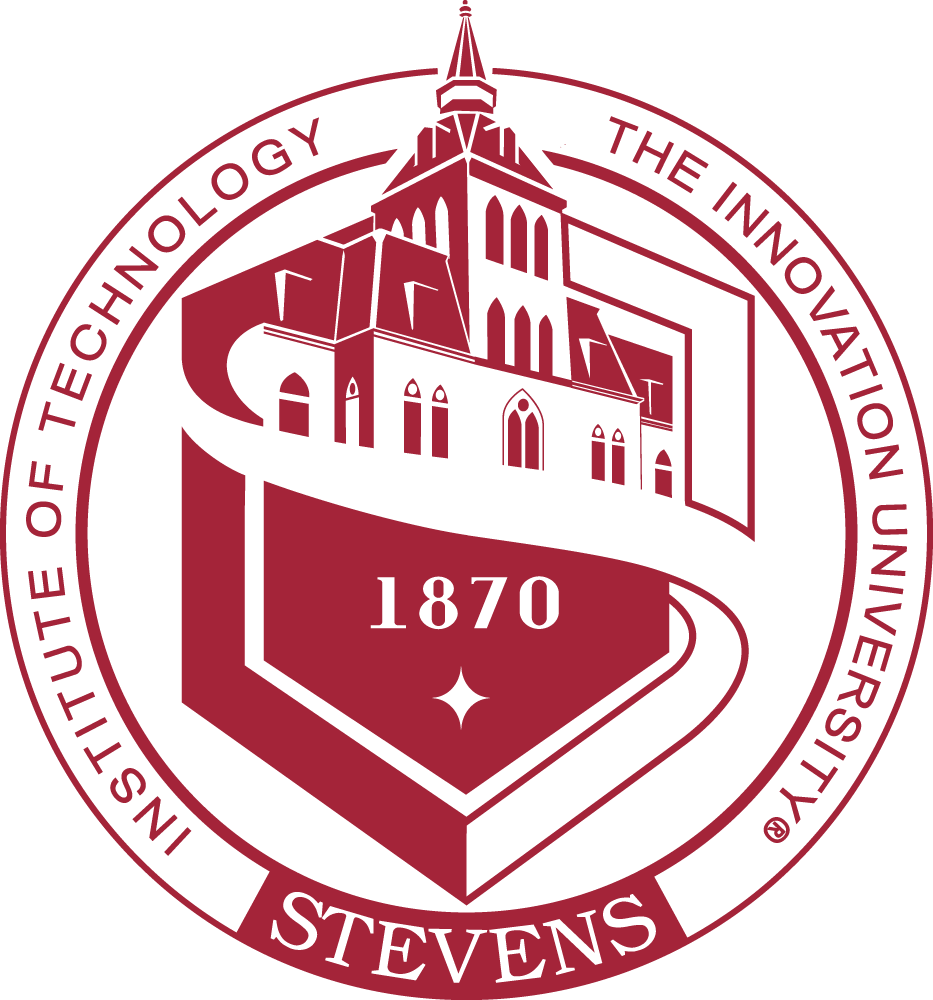}\\[1cm] % Include a department/university logo - this will require the graphicx package

\textsc{\LARGE Stevens Institute of Technology}
\\[0.5cm]
\textsc{\Large Master of Science in Financial Engineering}
\\[0.5cm]
\textsc{\large FE 800 - Special Research Problems}\\[3cm]

%----------------------------------------------------------------------------------------
%	TITLE SECTION
%----------------------------------------------------------------------------------------

\HRule \\[0.9cm]
{\huge \bfseries Learned Sectors}\\[0.4cm] % Title of your document
\textsc{\Large A Fundamentals-Driven Sector Reclassification Project}\\[0.5cm] % Minor heading such as course title
\HRule \\[1.5cm]

%----------------------------------------------------------------------------------------
%	DATE SECTION
%----------------------------------------------------------------------------------------

{\large May 16, 2019} % Date, change the \today to a set date if you want to be precise

%----------------------------------------------------------------------------------------
%	AUTHOR SECTION
%----------------------------------------------------------------------------------------

\vspace{\fill}

\begin{minipage}{0.4\textwidth}
\begin{flushleft} \large
\emph{Authors:}\\
\Large
Rukmal \textsc{Weerawarana} \\
Yiyi \textsc{Zhu} \\
Yuzhen \textsc{He}
\end{flushleft}
\end{minipage}
\hfill
\begin{minipage}{0.4\textwidth}
\begin{flushright} \large
\emph{Supervisors:} \\
\Large
Thomas \textsc{Lonon} \\
Ionut \textsc{Florescu} \\
Papa \textsc{Ndiaye} \\
Dragos \textsc{Bozdog}
\end{flushright}
\end{minipage}

\end{center}
\thispagestyle{empty}
\end{titlepage}

% Setting page numbering to roman
\newpage
\pagenumbering{roman}

% Abstact
\addcontentsline{toc}{chapter}{Abstract}

\section*{Abstract}
\markboth{Abstract}{}  % To create a phantom chapter "Abstract" for header formatting

Market sectors play a key role in the efficient flow of capital through the modern Global economy. Their use is widespread across the entire spectrum of market participants and observers, ranging from Governments using them to better regulate industry, to retail investors gaining exposure to particular segments of the economy through exchange traded funds tracking sector indices. We analyze existing sectorization heuristics, and observe that the most popular - the GICS (which informs the S\&P 500), and the NAICS (published by the U.S. Government) - are not entirely quantitatively driven, but rather appear to be highly subjective and rooted in dogma.

We examined alternative approaches to market sectorization, and found that returns-based methods were inherently flawed due to the significant bias of existing classifications on the structure of correlation distributions of the returns. Following this, we inspected determinants of firm value that would be intrinsically descriptive of the economic operating domain of a company. Building on inferences from analysis of the capital structure irrelevance principle and the Modigliani-Miller theoretic universe conditions, we postulate that corporation fundamentals - particularly those components specific to the Modigliani-Miller universe conditions - would be optimal descriptors of the true economic domain of operation of a company. Fundamentals data from Form 10-K for 15 features were downloaded for 362 companies in the S\&P 500, forming the feature space on which train our classification model.

To this end, we developed a new, objective data-driven sector classification heuristic, based on a HCA algorithm. We utilized this novel heuristic to generate a set of potential candidate learned sector universes, by varying the linkage method of the HCA algorithm (testing SLINK, CLINK, ALC, and WARD linkage methods), and the number of resulting sectors derived from the model (ranging from 5 to 19), resulting in a total of 60 candidate learned sector universes.

We then introduce \textit{reIndexer}, a backtest-driven sector universe evaluation research tool, to rank the candidate sector universes produced by our learned sector classification heuristic. \textit{reIndexer} backtests portfolios of synthetic exchange traded funds, constructed based on the specifications of a candidate sector universe. The backtest period was from January $1^\text{st}$ 2012 to December $31^\text{st}$ 2017, tracking the evolution of the portfolio daily. The backtest results of each classification universe are then evaluated against each other, to derive a de facto \textit{rank} for the candidate sector universes. This rank was utilized to identify the risk-adjusted return optimal learned sector universe as being the universe generated under CLINK (i.e. complete linkage), with 17 sectors.

Finally, we evaluate our risk-adjusted return optimal learned sector against the benchmark classification heuristic, the GICS S\&P 500 Classification. \textit{reIndexer} was used again to backtest the GICS classification universe against the optimal (complete linkage; 17 sectors) learned sector universe. We found that our learned sector universe portfolio outperformed the benchmark with respect to both absolute portfolio value, and the risk-adjusted return of the portfolio over the backtest period.

We conclude that we fully explored the scope of our thesis statement, and addressed our specific research goals through the successful development of a fundamentals-driven Learned Sector classification heuristic with a superior risk-diversification profile than the status quo classification heuristic.

\newpage

% Table of Contents
\tableofcontents
\newpage

% Setting page numbering to arabic
\clearemptydoublepage
\pagenumbering{arabic}

% This is where files are included, in order as necessary
% Sample include with the subfiles package:
% \subfile{sections/subfile_section}

% Introduction

\chapter{Introduction}

The United States today is home to approximately 20,000 publicly traded corporations. Despite only a small minority capturing the public eye on a regular basis, they all contribute to the foundation on which the modern Global economy is built. As postulated by nearly all economic theory, the efficient flow of capital and information through these markets is necessary for a healthy economy.

To this end, market sectors and the practice of sectorization have been an integral component of healthy markets, both in the United States and around the world. Market Sectors - in their ideal form - group together corporations of similar business function and economic operating arena for easier regulation, management, investment, etc. A related practice to that of market sectorization is Market Segmentation, the practice of dividing a market into subgroups of consumers (i.e. \textit{segments}).

The evolution of Market Segments and Market Sectors have historically been extremely useful metrics for gauging the development of the economy. There are four generally accepted stages of evolution in market segmentation; fragmentation, unification, segmentation, and hyper-segmentation.\citeFormat{\cite{Tedlow1996NewAmerica}} The United States economy - being the archetype on which this four-stage heuristic was built - developed through these four stages over the course of the last two centuries. Being a decidedly \textit{hyper-segmented} market today, there is a marked shift toward ever more narrow market segments. This shift has notably been amplified by the enable of hyper-targeted marketing and product delivery by technologies such as the smartphone.

\section{Applications of Market Sectors}

The United States Government began classifying companies into segments and sub-groups with the introduction of the Standard Industrial Classification (hereafter \textit{SIC}) system in 1937.\citeFormat{\cite{OfficeofStatisticalStandards-BureauoftheBudget1957HistoryClassification}} Certainly, the effective classification of companies is a key prerequisite to scalable monitoring and governance. The sheer scope of the modern economy guarantees the necessity of such classifications; the current scope of the economy spans - indisputably - all facets of Human culture. This gargantuan scope demands specialization, which - in turn - demands organization; motivating the need for market sectors.

Credit rating is the practice of evaluating the risk of a prospective counterparty in a transaction. This metric is integral to risk management, and relies on evaluating the probability that a candidate counterparty to a transaction will not default on their obligation. In addition to the idiosyncratic forces affecting any given corporation, its risks are often decomposed to market factors and - increasingly - market sector factors. This means that a positive outlook on a specific market sector would imply a more positive outlook for the constituent corporations composing that sector, underscoring the importance of appropriate and accurate sector assignment. The significance of accurate sector assignments is reaffirmed by the fact that all of the \textit{Big 3} credit rating agencies cite market sector rating as a key component in determining credit ratings.\citeFormat{\cite{StandardPoorsRatingServices2014CorporateMethodology}}\citeFormat{\cite{Hill2016FinancialMethodology}}\citeFormat{\cite{FitchRatings2019ProceduresRatings}}

\pagebreak

Finally, another major application of asset sectors is to provide investors with targeted exposure to specific segments of the the market. It is a well-known corollary of Modern Portfolio Theory that diversification provides savings an enhanced risk-return portfolio for any given basket of assets.\citeFormat{\cite{Markowitz1952PortfolioSelection}} This, combined with the excellent cost savings provided by modern Exchange Traded Funds (hereafter \textit{ETFs}), has led to a rapid prolification of these products in the Financial System today.

\section{Status Quo} \label{introduction:status_quo}

In the United States today, there are myriad sectorization taxonomies (hereafter \textit{sector universes}). To limit the scope of this analysis, we will focus on two of the three most popular sector classification systems\citeFormat{\cite{FidelityInvestments2019KnowIndustries}}; the GICS, and the ICB.

\subsection{GICS - Global Industry Classification Standard}

\begin{wrapfigure}[12]{r}{0.3\textwidth}
    \centering
    \vspace{\wrapfigadjustment}
    \fbox{
    \includegraphics[width=.9\linewidth]{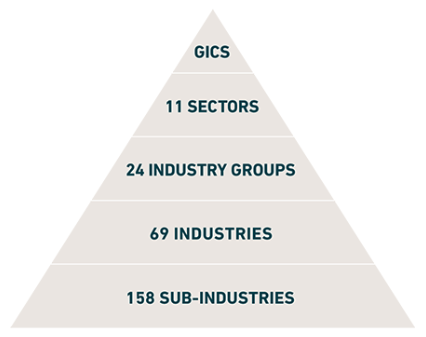}
    }
    \caption{Overview of the GICS sector classification universe.}
    \label{fig:introduction:gics_breakdown}
\end{wrapfigure}

The GICS (Global Industry Classification Standard)\citeFormat{\cite{MSCI-MorganStanleyCapitalInternational2019GICSMSCI}} is published by MSCI (Morgan Stanley Capital International) and S\&P (Standard \& Poor's). This classification is arguably the most widely used sector universe in the United States, and provides the basis for the popular S\&P 500 Market Sectors and Market Sector ETFs used in popular financial analysis resources.

Corporations are divided into four different categories, each in increasing order of specificity. The first classification is its \textit{sector} (the most general), followed by the \textit{industry group}, \textit{industry}, and finally \textit{sub-industry}, the most specific. The hierarchical taxonomy of these sector classifications are displayed in Figure~\ref{fig:introduction:gics_breakdown}\citeFormat{\cite{MSCI-MorganStanleyCapitalInternational2019GICSMSCI}}.

Additionally, the GICS methodology specification indicates that sector assignments and assignment updates are made primarily based on three factors; the primary source of revenue, earnings and market perception, and finally - in the case of a new company - information derived from the company prospectus.\citeFormat{\cite{SPGlobalMarketIntelligence2018GlobalMethodology}}

\textit{Note:} A key change in this sector classification taxonomy was made in the Fall of last year (September 2018).\citeFormat{\cite{MSCIResearch2018ConsultationIndexes}} Specifically, the previously-labeled \textit{Telecommunications Services} sector was broadened and renamed to \textit{Communication Services}. Company sector assignment changes were also made commensurate to the name and scope change:

\begin{itemize}
    \item Media companies were moved from \textit{Consumer Discretionary} to \textit{Communication Services}
    \item Internet services companies were moved from \textit{Information Technology} to \textit{Communication Services}
    \item E-Commerce companies were moved from \textit{Information Technology} to \textit{Consumer Discretionary}
\end{itemize}

\subsection{ICB - Industry Classification Benchmark}

The ICB (Industry Classification Benchmark)\citeFormat{\cite{FTSEInternationalLimited2019IndustryRussell}} is published by FTSE International (previously jointly owned by Dow Jones and FTSE). Similarly to the GICS, the ICB also classifies corporations into four increasingly specific categories; \textit{industry} (the most general), \textit{supersector}, \textit{sector}, and finally \textit{subsector}, the most specific. A visualization of the ICB taxonomy is reproduced in Figure~\ref{fig:introduction:icb_breakdown}\citeFormat{\cite{FTSEInternationalLimited2019IndustryRussell}}.

Similar to the GICS, ICB too utilizes three main criteria when classifying companies into specific sectors, and other categories. They are; the primary source of revenue, description in annual filings, and - in the case of a new company - information derived from the company prospectus, or regulatory filing descriptions (company-supplied).

\pagebreak

\section{Key Limitations} \label{introduction:key_limitations}

\begin{wrapfigure}[11]{R}{0.3\textwidth}
    \centering
    \vspace{\wrapfigadjustment}
    \fbox{
    \includegraphics[width=.9\linewidth]{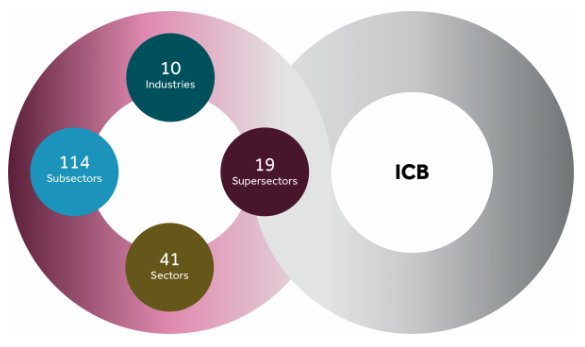}
    }
    \caption{Overview of the MSCI sector classification universe.}
    \label{fig:introduction:icb_breakdown}
\end{wrapfigure}

In this section, we analyze the GICS and ICB sector classification schemes described above through the lens of their limitations. We then utilize these key limitations to inform our research goals.

As discussed above, both GICS and ICB utilize information from company prospectuses to determine an initial sector assignment. Unfortunately, this inherently implies that the initial classification is not based on an objective criteria, and is highly subject to the initial vision of the authors of the company prospectus. This is in stark contrast to utilizing a company-specific quantifiable metric, and relies on accurate reporting in the initial prospectus; a document whose authors are highly incentivized to inflate in grandiosity.

Furthermore, the initial sector groups and constituent sectors in both classification schemes are defined based on a qualitative analysis of the economy, as opposed to a quantitatively-driven process. Similarly, the number of sectors in the market is also arbitrary, and not based on a quantifiable or objective metric.

Additionally, as implied by the reorganization of the GICS classification scheme in 2018, there appears to be no strong objective criteria governing the creation and deletion of new sectors. The companies reassigned during this reorganization were not new, and imply that the underlying newly created sector existed before its recognition by the GICS scheme.

As evidenced by the short list of limitations outlined in this section, the status quo of market sector classifications are far from perfect. To this end, we hope to develop a new \textit{Learned Sectors} scheme, addressing each of the limitations described above.

% Research Goals

\chapter{Research Goals} \label{research_goals}
    
In this section, we outline our overarching thesis statement. Additionally, we also isolate specific research goals based on this thesis statement, which we will address in sequence throughout the report. Through tackling each of the stated research goals, we hope to address the full scope of our thesis statement.

\section{Thesis Statement} \label{research_goals:thesis_statement}

% See: https://www.slideshare.net/linjaaho/how-to-make-boxed-text-with-latex
% See: ftp://ftp.dante.de/tex-archive/graphics/bclogo/doc/bclogo-doc.pdf [Its in French lmfao]
\begin{center}
    \begin{minipage}{0.7\textwidth}
        \begin{bclogo}[couleur=blue!30, arrondi=0.1, logo=\bcloupe, ombre=false]{\;Thesis Statement}
            Utilize relationships in the idiosyncratic characteristics of corporations to inform a fundamentals-driven, non-subjective sector classification framework.
        \end{bclogo}
    \end{minipage}
\end{center}

\vspace{1em}

The thesis statement above encapsulates - at a very high level - the key issues encountered in existing sector classification heuristics, and how we plan to address these limitations. We believe that - given the objectivity of our classification - our \textit{Learned Sectors} will provide a better basis for natural economic diversification. This is due to the fact that the underlying division of sectors, and assignment of corporations into those sectors will be objectively and quantitatively driven, rather than subjectively and qualitatively driven as is the status quo.

To combat the primary issue of the previously discussed classification heuristics (their lack of objectivity, particularly for newly classified companies), we seek to restrict our input data to the classification algorithm to reduce complexity. We believe that this restriction sufficiently limits the scope of our investigation, while also providing us with a quantifiable, objective measure of a capital structure, which - we believe - will reflect underlying economic function.

Given the constraint on our input data - motivated by end goal of increasing objectivity and reducing subjective involvement in the classification of the corporations - we also postulate that we will use data driven algorithms to derive potential classifications. That is, we plan to use entirely Unsupervised Learning methods, which do not require the definition of a \textit{cost function}. This lack of a cost function - in addition to reducing complexity of the project - also removes another aspect of potential bias in the classification of the companies.

However, a shortcoming of this approach is that we will have a clustering algorithm parameterized by some set of arbitrary parameters, which will map a set of potential corporations to a set of potential market sectors. In keeping with the spirit of objectivity, we cannot arbitrarily assign values to the parameters, and thus must derive a method for \textit{ranking} our potential sector universes. Note that this ranking cannot be on an objective scale, but rather would be a relative ranking comparing each candidate universe to its peers, thus maintaining objectivity of the ranking.

Finally, we hope to evaluate our sector classification against a benchmark sector universe; the \textit{GICS S\&P 500 Sector Classification} (hereafter \textit{the benchmark}). To do this, we will use the same metric(s) utilized in ranking the candidate sector universes, and maintain any specific methodology used to compute those rankings between the \textit{best} Learned Sector Universe, and the benchmark. If our initial hypothesis is correct, the superior risk diversification benefit inherent to our fundamentals-driven sector divisions will lead our Learned Sector Universe to outperform the benchmark with respect to the evaluation metric.

\pagebreak

\section{Specific Research Goals} \label{research_goals:specific_research_goals}

Here, we encapsulate the gist of the previous discussion of our thesis statement in a collection of specific research goals. We will then address each of these research goals in sequence through the rest of the report, thereby fully exploring the scope of our thesis statement in the process.

\vspace{2em}

\begin{table}[h!]
    \centering
    \begin{tabular}{| c | c |}
        \hline
        & \\
        \textbf{Number} & \textbf{Description} \\
        & \\
        \hline
        & \\
        RG-1 & Utilize data-driven algorithms to derive a truly objective classification heuristic. \\
        & \\
        \hline
        & \\
        RG-2 & Rank candidate sector universes against each other using entirely objective criteria. \\
        & \\
        \hline
        & \\
        RG-3 & Evaluate our risk-adjusted return optimal sector universe against the benchmark. \\
        & \\
        \hline
    \end{tabular}
    \caption{Specific, itemized research goals of the Learned Sectors project.}
    \label{table:research_goals:research_goals}
\end{table}

% Literature Review

\chapter{Literature Review} \label{literature_review}

Having identified concrete research goals designed to fully explore the scope our thesis statement, we explored existing research in the field of market segmentation.

Due to their many applications and widespread use across the world, the problem of market sectorization has been approached through myriad lenses. For example, some authors have focused on sub-dividing an already established (i.e. dogmatic) sectors, while others have focused on the specific learning algorithms that may be successfully applied to the task of hierarchical decomposition of a set of related entities.

To best navigate the large corpus of research that is relevant to our research goals, we divide the presentation of our literature review into three sections:

\begin{itemize}
    \item \textbf{Existing Heuristic Evaluation}: Evaluating the existing dominant sectorization heuristics.
    \item \textbf{Alternative Approaches to Market Sectorization}: Exploration of unorthodox approaches to market sectorization.
    \item \textbf{Relationship between Economic Sectors and Fundamentals Data}: Analysis of the relationship between company fundamentals data and their business function.
\end{itemize}

\section{Existing Heuristic Evaluation} \label{literature_review:existing_heuristic_evaluation}

Originally established in the United States in 1937, the SIC\citeFormat{\cite{OfficeofStatisticalStandards-BureauoftheBudget1957HistoryClassification}} is a system for classifying industries with a four-digit code. Due to its abundant use in industry, the SIC Classification system has been widely used as an instrument in published Finance and Accounting Research. In 1997, the North American Industry Classification System\citeFormat{\cite{UnitedStatesOfficeofManagementandBudget1997NorthNotice}} (hereafter \textit{NAICS}) was been adopted as an alternative to the SIC by various Government agencies, and is often cited interchangeably with the SIC in certain research. The key difference between the two heuristics is that NAICS is production-oriented, whereas the SIC is market-oriented.\citeFormat{\cite{EconomicClassificationPolicyCommitteeECPC20072017Manual}}

In \textit{The Impact of Industry Classification Schemes on Financial Research},\citeFormat{\cite{Weiner2005TheResearch}} the author evaluates the usage of existing classification heuristics in Finance and Accounting Research published in major research journals. The author finds that approximately 30\% of all research published in the top 3 Finance, and top 2 Accounting journals utilize industry classification systems. Given this relatively abundant usage, it is extremely concerning that the underlying heuristic itself is not entirely objective or quantitatively derived, as discussed in Section~\ref{introduction:key_limitations}.

They are mainly used for sample restriction (34\%), comparable company selection (31\%), and detection of industry effects (12\%). Under the reasonable assumption that Finance and Accounting Research is utilized when publishers create or update classification heuristics, this behavior of widespread use in existing work may be indicative of a feedback pattern, where existing structural dogmas of prior classification heuristics are implicitly reimposed on new systems. Additionally, the author also discovers that approximately 45\% of all corporations change their industry over time based on the SIC Classification, and 20\% based on the GICS\citeFormat{\cite{MSCI-MorganStanleyCapitalInternational2019GICSMSCI}} industry classifications. This result highlights the lack of temporal stability prevalent in popular classification heuristics.

Despite this apparent lack of temporal stability of assignment, in her paper \textit{Structural change and industrial classification},\citeFormat{\cite{Hicks2011StructuralClassification}} the author evaluates the impact of the slow rate of change of existing heuristics with respect to the addition and deletion of new and emergent industry groups to sector classification taxonomies. The author recognizes the fact that existing heuristics provide an incalculable resource to researchers. Due to this, she also infers that the co-dependence of researchers and classification publication agencies have led to existing classification schemes becoming de facto descriptors of economic industries, as opposed to the other way around. The author then performs an empirical analysis of the classification of highly innovative firms providing products and services in gaming devices, packaging, filtration, photonics, imaging, biomedical research, and fabless semiconductor design. Through her analysis, she finds significant vertical disintegration in existing classification heuristics.

The performance of NAICS and the GICS S\&P 500 Classification heuristics are evaluated from a quantitative perspective in \textit{A comparison of industry classification schemes: A large sample study}.\citeFormat{\cite{Hrazdil2013AStudy}} The authors perform individual linear regressions of a selection of fundamentals and earnings data of companies in the S\&P Composite 1500 index against the sector assignments implied by the NAICS and GICS heuristics (among others). The various linear regressions are compared through the lens of an adjusted $R^2$-derived metric. The results indicate that the GICS heuristic performed best, but the maximum adjusted $R^2$-derived metric (realized on the monthly returns vs. GICS sector linear regression) was only 13.59\%. Clearly, this is an extremely sub-optimal result.

\section{Alternative Approaches to Market Sectorization}

While not directly applied to the specific research problem of Market Sectorization, \textit{Correlation Structure and Evolution of World Stock Markets: Evidence from Pearson and Partial Correlation-Based Networks}\citeFormat{\cite{Wang2018CorrelationNetworks}} provides excellent insight into the correlation structure of returns in the more generalized global economic environment. The authors analyzed daily price indices of 57 stock markets from 2005 to 2014, and inspected the distributions of the Pearson and Partial correlations between pairs of stock markets. In addition to affirming Economic theory through the confirmation that correlations between markets increase substantially during crisis, they also found that large groups of correlated markets exist based on their geographic location.

The authors' results confirm that the existence of pre-determined groupings of assets significantly affects the correlation distribution of those assets over time. Treating the geographic location of markets as a proxy for a generalized pre-existing group, we can extrapolate these effects to the more localized United States market. This generalization suggests that the existing sector groups would have a significant effect on the historical returns of companies in a given sector. This in turn implies that the usage of historical asset returns would introduce bias from existing sector groups to a new heuristic.

\textit{Marketing segmentation using support vector clustering}\citeFormat{\cite{Huang2007MarketingClustering}} explores the application of a support vector clustering (a permutation of the support vector machine) to a relatively low-dimensional marketing dataset to derive clusters. The support vector clustering method is parameterized with a cluster count, and a random initialization of cluster centroids. Additionally, support vector clustering does not guarantee cluster assignment for all data points, and outliers remain unclassified. This approach is then compared to a $K$-means clustering and self-organizing feature map (SOFM) method, and is found to perform better based on a mean and standard error index evaluation. Despite appearing to be a promising approach in the authors' sample case study, the support vector clustering algorithm's cluster count parameterization, and its treatment of outliers do not make it suitable for the problem of market sectorization.

The authors of \textit{A purchase-based market segmentation methodology}\citeFormat{\cite{Tsai2004AMethodology}} apply a genetic algorithm to cluster transactional purchase data from a set of customers, with the end goal of training an RFM (Recency, Frequency, Monetary Value) model. The genetic algorithm, along with a cost function to assess the fidelity of fit, is used to segment customers into unique clusters based on their purchasing data. The iterative and stochastic behavior of the genetic algorithm ensures that the resulting cluster assignments are extremely stable, while also being non-variant with respect to centroid initialization. However, as with the previously discussed support vector clustering method, this approach is hindered by the required prior specification of a cluster count, as well as not being hierarchical in nature.

\pagebreak

\section{Relationship between Economic Sectors and Fundamentals Data} \label{literature_review:economic_sectors_fundamentals}

\textit{The determinants of capital structure in transitional economies}\citeFormat{\cite{Delcoure2007TheEconomies}} provides an in-depth quantitative analysis of the alignment of traditional optimal capital-structure dogma against the real-world behavior of companies in transitional economies. The results suggest that while some traditional capital structure theories are indeed applicable to transitional economies, a large portion of capital structures are not well described by these traditional theories. Rather, the author finds that disparities in legal systems, shareholder power and demographics, and corporate governance provide a significantly better frame of explanation for the variance observed in capital structure.

\textit{Determinants of capital structure of Chinese-listed companies}\citeFormat{\cite{Chen2004DeterminantsCompanies}} analyzes the capital structures of corporations in China, providing a much better proxy for the large developed United States economy. The results presented by the authors echo that of \citeauthor{Delcoure2007TheEconomies}, asserting that traditional theory does not fully describe the distribution of capital structures in China. Furthermore, the authors also allude to myriad other factors affecting capital structure, similar to \citeauthor{Delcoure2007TheEconomies}. This work confirms that the dynamics of the determinants of capital structures observed in transitional economies are portable to larger, more established economies.

As highlighted above in Section~\ref{literature_review:existing_heuristic_evaluation}, existing sectorization heuristics do not exhibit strong temporal stability. Thus, to avoid overfitting against the changing dynamics of a market when designing our new sectorization heuristic, we postulate that it would be beneficial to treat the market as constantly transitional. Under this assumption, the findings of \citeauthor{Delcoure2007TheEconomies} can be applied to our prospective heuristic to great effect. The author's findings would suggest that we focus on determinants of the factors listed above to best capture the idiosyncratic dynamics of a given company, as opposed to focusing on traditional metrics of performance, such as asset returns.

Under the assumptions of the Modigliani-Miller theoretic universe (no taxes, bankruptcy costs, agency costs, and asymmetric information), the capital structure irrelevance principle\citeFormat{\cite{Modigliani1958TheInvestment}} postulates that - in an efficient market - the value of a firm is unaffected by how that firm is financed. However, given that all of the conditions of the theoretic universe are violated in the real world, this leads to the profound realization that capital structure is the single most important determinant of firm value.\citeFormat{\cite{Vernimmen2005CorporatePractice}} Based on the observation that firm value is derived from a company's intrinsic economic domain of operation, the capital structure irrelevance principle - in conjunction with the violation of Modigliani-Miller universe assumptions - implies that capital structure is governed by the idiosyncrasies of the true economic domain of a company.

Given that the exact quantified magnitude of violation of each of the Modigliani-Miller theoretic universe conditions are inherently specific to a given economic segment, we postulate that corporation fundamentals reflective of capital structure - particularly those components specific to the Modigliani-Miller universe conditions - would be the optimal descriptors of the true underlying economic domain of a company.

% Model Data

\chapter{Model Data} \label{model_data}

In this section, we describe the data sources used in our project, and identify specific features to be used in our sector classification heuristic. Additionally, we also describe the benchmark sector classification universe that we will use to evaluate our final results. This section begins the discussion of our first research goal, RG-1.
  
 \begin{table}[h!]
    \centering
    \begin{tabular}{| c | c |}
        \hline
        &  \\
        RG-1 & Utilize data-driven algorithms to derive a truly objective classification heuristic. \\
        & \\
        \hline
    \end{tabular}
\end{table}

\section{Fundamentals Data Overview}

In the previous section (see page~\pageref{literature_review:economic_sectors_fundamentals}), we explored the effect of the violation of the Modigliani-Miller theoretic universe conditions on the capital structure irrelevance principle, in conjunction with observations of the dynamics of the determinants of capital structure in transitional and established economies. The logical corollary of this analysis is that fundamentals data reflective of capital structure - particularly those specific to the Modigliani-Miller universe conditions - are optimal descriptors of the economic domain of a company.

Based on this conclusion, we identified earnings data from Form 10-K\citeFormat{\cite{U.S.SecuritiesandExchangeCommission2019Form10-K}} filings to be our model input data. This data was retrieved for 362 companies in the S\&P 500 Index, for every year from 2010 to 2017, from the Compustat Database\citeFormat{\cite{SPGlobalMarketIntelligence2019CompustatIntelligence}}, via the Wharton Research Data Services\citeFormat{\cite{TheWhartonSchool1993WhartonServices}} Cloud (hereafter \textit{WRDS}).

\section{Feature Selection}

Given the variability of earnings reports, we identified 15 specific features from the annual Balance Sheet, Income Statement, and Statement of Cash Flows guaranteed to exist for all companies in our dataset. In addition to being common across all companies, they were also isolated on the basis of being related to, or direct arguments of, the capital structure of the company.

\begin{table}[h]
    \centering
    \begin{tabular}{|c|c|c|}
        \hline
        Total Assets & Cash \& Equivalents & Receivables \\
        \hline
        Inventories & Sales & Cost of Goods Sold \\
        \hline
        Gross Profit & Operating Cash Flow & Operating Income \\
        \hline
        Depreciation, Depletion \& Amortization & Interest Expense & Non-Operating Income/Expense \\
        \hline
        Income Taxes & Advertising Expense & Research \& Development Expense \\
        \hline
    \end{tabular}
    \caption{Selected model input data features from Form 10-K for each company.}
    \label{table:model_data:features}
\end{table}

\pagebreak

\begin{figure}[h]
    \centering
    \fbox{
    \includegraphics[width=.9\linewidth]{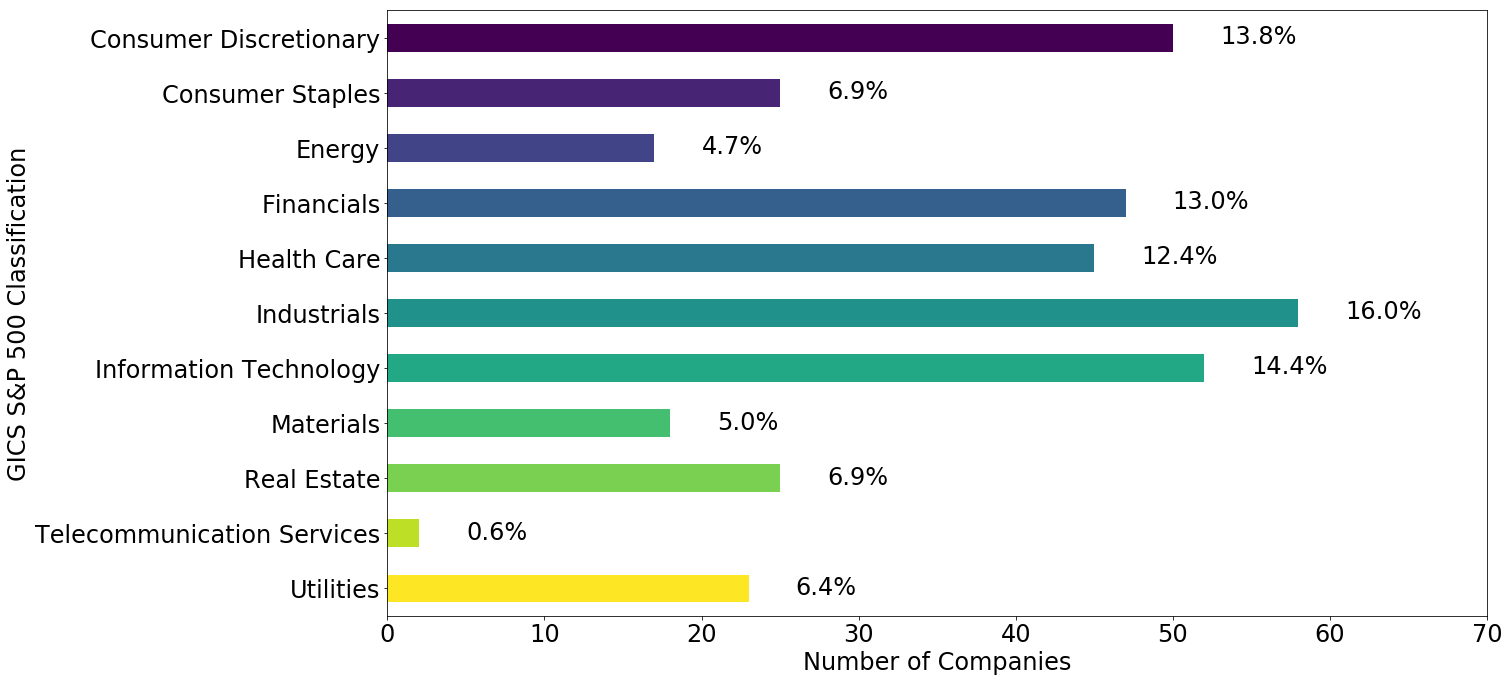}
    }
    \caption{Distribution of input data companies ($n = 362$) across sectors in the benchmark universe (i.e. GICS S\&P 500 Classification).}
    \label{fig:model_data:sp500_sector_distribution}
\end{figure}

\section{Benchmark Sector Universe}

To evaluate our final learned sector universe and fully address RG-3, we identified the GICS S\&P 500 Classification\citeFormat{\cite{MSCI-MorganStanleyCapitalInternational2019GICSMSCI}} (hereafter \textit{benchmark universe}) to be our benchmark. Unfortunately, the complete dataset of sector assignments our benchmark universe is proprietary. Due to this, we were unable to collate historical sector assignments, and were limited to the latest sector assignments for companies in our input data space.

Due to the disparity in temporal alignment between our data, we decided to utilize only the latest available data for our learned sector universe evolutions. That is - unless stated otherwise - we only utilized learned sector assignments implied by the 2017 10-K Form data for the remainder of this project.

The distribution of the 362 companies in our input data across various sectors in the benchmark universe is displayed in Figure~\ref{fig:model_data:sp500_sector_distribution}.

% Unsupervised Learning Methods Survey

\chapter{Learning Methods Survey} \label{learning_methods_survey}

In this chapter, we outline a set of ideal characteristics and desired behavior of an ideal candidate classification algorithm, and conduct a survey of potential unsupervised learning methods. We then evaluate each of these surveyed methods against our selection criteria, and determine the optimal method with which to proceed.

\section{Evaluation Criteria}

Despite being not entirely objective, existing classification heuristics have a range of desirable behavior that we would wish to replicate with our candidate clustering algorithm. Additionally, we would want to replicate this behavior while also maintaining objectivity and stability in our new heuristic by utilizing a highly nonparametric learning method.

In particular, we would be extremely interested in preserving the nested hierarchical clustering behavior of the current schemes. That is, to be able to classify a market into sectors, and in turn those sectors into subsectors. Furthermore, it would be desirable to be able to determine these nested sub-sectors in the context of the greater market, rather than in isolated analysis of a particular sector.

Additionally, we would also like to vary the number of resulting sectors of our algorithm while maintaining stability. That is, if we were to request two sectors from a heuristic that was initially resulting in four sectors, the two sectors would be some combination of the initial four sectors, as opposed to an entirely new segmentation profile. This behavior is reflective of the real world, where economic sectors often exhibit nesting, as opposed to independent clustering.

Finally, as per RG-1 (see Section~\ref{research_goals:specific_research_goals}), we are extremely motivated to design a heuristic that is either entirely non-parametric, or parameterized with highly objective, quantitatively derived criteria. In addition to preserving mathematical objectivity of our results, a nonparametric approach would ensure that no personal biases - either explicit or implicit - are introduced to the final learned sectors.

\section{Candidate Learning Methods}

Given the required behavior outlined above, we evaluated three major families of clustering algorithms. We empirically evaluate each clustering technique through the lens of the requirements outlined above.

\subsection{$K$-means Clustering}

$K$-Means Clustering is a method of partitioning $n$-dimensional data into a set of $K$ distinct clusters. The basic algorithm is outlined below\citeFormat{\cite{Lloyd1982LeastPCM}}:

\begin{gather*}
    \text{Let $C_1, C_2, \ldots, C_K$} = \text{Set of K possible clusters} \\
    \text{Let $W(C_k)$} = \text{Measure of pariwise difference of observations in a cluster} \\
    \text{Let $x_{ij}$} = \text{$j^\text{th}$ feature in cluster $i$ with coordinates $x$} \\
    \Rightarrow W(C_k) = \frac{1}{|C_k|} \sum_{i, i^\prime \in C_k} \sum_{j = 1}^p (x_{ij} - x_{i^\prime j})^2 \\
    \\
    \Rightarrow \text{$K$-Means Clusters}
    = \underset{C_1, \ldots, C_k}{\text{minimize}} \sum_{k=1}^K W(C_k)
    = \underset{C_1, \ldots, C_k}{\text{minimize}} \sum_{k=1}^K \frac{1}{|C_k|} \sum_{i, i^\prime \in C_k} \sum_{j = 1}^p (x_{ij} - x_{i^\prime j})^2
\end{gather*}

Notice that in the algorithm outlined above, the $K$-means clustering process requires two sets of parameters at initialization. First, it requires the number of target clusters, $K$, as well as a set of random initializations for cluster centroids, $\frac{1}{|C_k} \sum_{i \in C_k} x_{ij}$. This high level of parameterization, coupled with the clear lack of congruity of assignment across varying values of $K$ make this family of algorithms poorly suited to the task of sector classification, as per the constraints detailed above.

\subsection{Support Vector Classifier}

The support vector classifier is based on the notion of finding a set of hyperplanes in a higher dimensional feature space that optimally divides a set of data into classes. Data is mapped to a higher dimensional space to ensure orthogonal hyperplanes in the divisions of the clusters. The support vector classifier objective function is outlined below\citeFormat{\cite{Ben-Hur2001SupportClustering}}:

\begin{equation*}
    \begin{aligned}
        & \underset{R, a, \boldsymbol{\alpha}}{\text{minimize}} & & R^2 - \sum_i \alpha_i (R^2 - ||x_i - a||^2) \\
        & \text{subject to}
        & & \alpha_i \geq 0 \\
        & & & (R^2 - ||x_i - a||^2) = 0 \; \forall \; i \; \text{(KKT Condition)}
    \end{aligned}
\end{equation*}

As indicated by the literature, and through inspection of the objective function, it is clear that the support vector classifier is parameterized on the kernel used for optimization, as well as the specific loss function employed during model training. Furthermore, this model also optimizes to a fixed numbed of sectors, as opposed to a dynamic number. Therefore, this method too is inappropriate as per the evaluation criteria.

\subsection{Hierarchical Cluster Analysis} \label{learning_methods_survey:hca}

Hierarchical clustering is a greedy algorithm which seeks to build clusters following either an agglomerative, or a divisive approach.\citeFormat{\cite{Ward1963HierarchicalFunction}} Agglomerative clustering is \textit{bottom-up}, with each observation starting it its own cluster, whereas divisive is \textit{top-down}, with all observations starting in one cluster and splits performed recursively at each level. The clusters output by this algorithm are determined by two model settings; the distance metric (i.e. the algorithm for computation of pairwise distance between observations), and the linkage method, which specifies the algorithm governing the dissimilarity of entire sets, as a function of the pairwise distances of observations in those sets.

This method has the distinct advantage of being entirely additively hierarchical, with groups being nested as described in the evaluation criteria. Furthermore, this method is entirely nonparametric, with the sole exception being the choice of linkage and distance metric. Additionally, it is extremely stable with varying sector counts. This is a direct result of the greedy nature of the algorithm, as it does not recompute the hierarchy each time a new cluster arity is extracted, but rather just changes the level of extraction from the same hierarchy.

As per the evaluation of the different families of learning methods detailed in this chapter, we selected Hierarchical Clustering to be the basis of our Learned Sectors classification heuristic.

% Hierarchical Clustering Model

\chapter{Hierarchical Clustering Model} \label{hierarchical_clustering_model}

% Quick overview of HCA again
% Distance metric

% Talk about maintaining "unsupervised" nature; build search space now evaluate next
% Criteria being varied in search space; linkage [4], number of sectors [15; range from 5-19]
% Visualize search space

In the previous chapter, we evaluated major families of learning methods and identified the Hierarchical Clustering Analysis (hereafter \textit{HCA}) algorithm to be the method best aligned with the research goals of the project. Here we outline the specifics of our approach to applying HCA to our model input data, and build a search space of candidate universes to be evaluated, fully addressing RG-1.

\section{HCA Overview}

As outlined in Section~\ref{learning_methods_survey:hca}, hierarchical clustering is a greedy learning algorithm which seeks to construct a hierarchy of clusters. The greedy nature of this algorithm results in extremely high computational complexity for any given model fit, but is extremely stable in its solution. Furthermore, it is an $\mathcal{O}(1)$ complexity operation to extract classifications of varying arity due to the persistent hierarchical nature of the algorithm.

One of the main requirements of our heuristic is the ability to create sector universes with varying numbers of sectors. Because of this, we elected to utilize an Agglomorative approach to clustering. That is, we utilize a \textit{bottom-up} HCA model, where each company begins in its own sector, with larger clusters derived at each successive step of the tree by merging existing pairs of clusters.

Any given HCA algorithm tree is parameterized by two distinct settings; the distance metric, and the linkage method. To best understand the potential candidate universes that may be generated by this HCA-driven classification heuristic, we analyzed each of these model settings in turn:

\subsection{Distance Metric}

The distance metric is the measure of the distance between pairs of observations. This setting primarily affects the shape of the clusters. Due to the fact that our model input data is exclusively in monetary units (i.e. United States Dollars), we do not intend to transform the existing metric of wealth reflected by the dollar value measurement. Thus, we chose to use the $\ell^2$ (i.e. Euclidean) distance metric for our heuristic.

\begin{gather*}
    \text{Let $\boldsymbol{p}, \boldsymbol{q}$} = \text{Cartesian coordinates $\boldsymbol{p} = (p_1, \ldots, p_n)$ and $\boldsymbol{q} = (q_1, \ldots, q_n)$ where $\{\boldsymbol{p}, \boldsymbol{q}\} \in \mathbb{R}^{n \times 2}$} \\
    \text{Let $dist(\boldsymbol{p},\boldsymbol{q})$} = \text{$\ell^2$ (i.e. Euclidean) distance between points $\boldsymbol{p}$ and $\boldsymbol{q}$} \\
    \\
    \Rightarrow dist(\boldsymbol{p}, \boldsymbol{q}) = dist(\boldsymbol{q}, \boldsymbol{p})
    = \sqrt{(p_1 - q_1)^2 + (p_2 - q_2)^2 + \cdots + (p_n - q_n)^2}
    = \sqrt{\sum_{i=1}^n (p_i - q_i)^2}
\end{gather*}

\subsection{Linkage Method}

The second setting governing the behavior of the HCA algorithm is the selection of a linkage method. The linkage is a measure of distance between sets of observations as a function of the pairwise distances between observations. There are four major linkage method choices that we evaluate in our HCA model:

    $$ \text{Let $A, B, C, X, Y$} = \text{Sets (i.e. clusters) of observations} $$
    $$ \text{Let $C$} = X \cup Y $$
    $$ \text{Let $N$} = |A| + |X| + |Y|, \; \text{where} \; |\alpha| = \text{Cardinality}(\alpha) $$
    
\hspace{7em} \textbf{Single Linkage\citeFormat{\cite{Sibson1973SLINK:Method}}:}
        $$ d_\text{SLINK}(A, B) = \text{min} \; dist(\boldsymbol{a}, \boldsymbol{b})  \; \forall \; \{ \boldsymbol{a}, \boldsymbol{b} : \boldsymbol{a} \in A, \boldsymbol{b} \in B \} $$
    
\hspace{7em} \textbf{Complete Linkage\citeFormat{\cite{Defays1977AnMethod}}:}
        $$ d_\text{CLINK}(A, B) = \text{max} \; dist(\boldsymbol{a}, \boldsymbol{b})  \; \forall \; \{ \boldsymbol{a}, \boldsymbol{b} : \boldsymbol{a} \in A, \boldsymbol{b} \in B \} $$
    
\hspace{7em} \textbf{Average Linkage\citeFormat{\cite{Seifoddini1989SingleApplications}}:}
        $$ d_\text{ALC}(A, B) = \frac{1}{|A| \cdot |B|} \sum_{\boldsymbol{a} \in A} \sum_{\boldsymbol{b} \in B} dist(\boldsymbol{a}, \boldsymbol{b}) $$
    
\hspace{7em} \textbf{Ward (variance minimization) Linkage\citeFormat{\cite{Ward1963HierarchicalFunction}}:}
        $$ d_\text{WARD}(C, A) = \sqrt{ \frac{|A| + |X|}{N} d_\text{WARD}(A, X)^2 + \frac{|A| + |Y|}{N} d_\text{WARD}(A, Y)^2 - \frac{|A|}{N} d_\text{WARD}(X, Y)^2 } $$

\section{Unsupervised Learning Approach}

\begin{wrapfigure}[14]{r}{0.45\textwidth}
    \centering
    \vspace{\wrapfigadjustment}
    \fbox{
    \includegraphics[width=.9\linewidth]{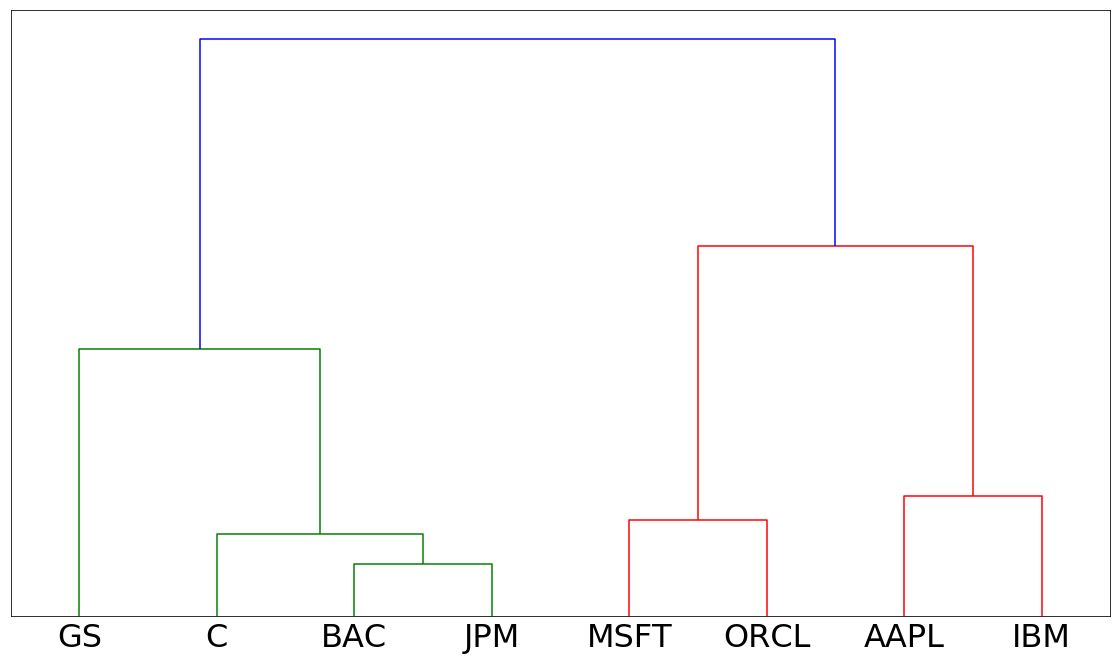}
    }
    \caption{Dendrogram of a sample hierarchical clustering model result.}
    \label{fig:hierarchical_clustering_model:sample_dendogram}
\end{wrapfigure}

As per RG-1 (see Section~\ref{research_goals:specific_research_goals}), we seek to create an entirely objective classification heuristic. Logically, this implies that we utilize an entirely nonparametric approach when designing the classification heuristic. However, as discussed above, the HCA algorithm is parameterized by both the distance metric, and the linkage method (in addition to a posterior selection of the number of sectors).

To work around the semi-supervised nature of the learning method, we elected to utilize HCA to build a search space of potential candidate sector universes. Following this, we will address our second research goal, RG-2, to rank these candidate sector universes against each other to determine the optimal learned sector classification.

Note that this search-space generation varies the sector count and linkage method parameters of the HCA model, but not the distance metric. This is due to the fact that we wish to preserve the monotonic and geometric difference of magnitudes of wealth implied by the dollar values of our input data.

Figure~\ref{fig:hierarchical_clustering_model:sample_dendogram} is a dendrogram of a sample HCA model generated on a subset of the model input data. Despite having a very high time complexity for model training, the HCA model thrives in its ability to extract sector classifications of varying arity from a HCA model. Its ability to perform this action in constant time complexity greatly enhanced our ability to generate a large search space of candidate learned sector universes.

\pagebreak

\section{Learned Sector Universe Search Space}

\begin{figure}[h]
    \centering
    \fbox{
    \includegraphics[width=.8\linewidth]{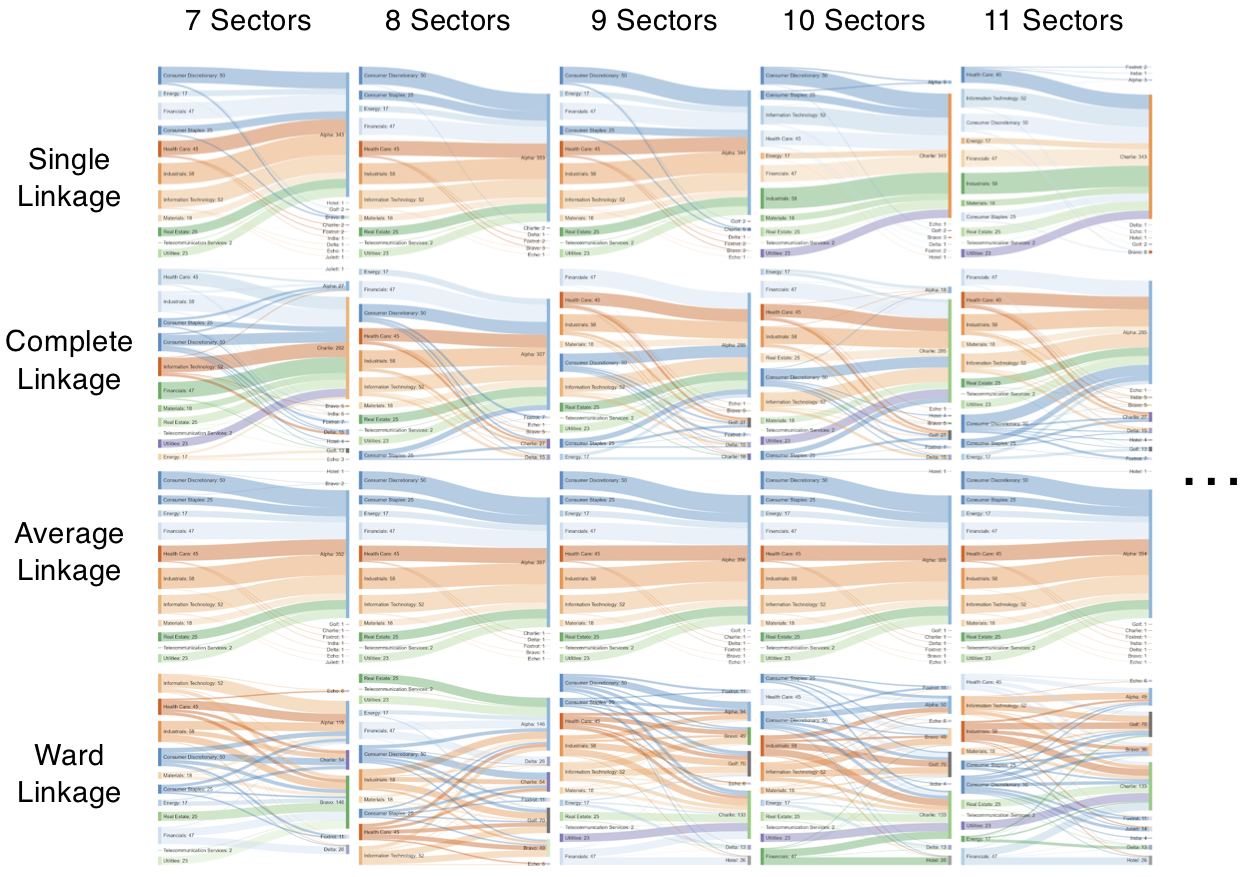}
    }
    \caption{Candidate learned sectors partial search space visualization.}
    \label{fig:hierarchical_clustering_model:partial_search_space}
\end{figure}

With the end goal of building a comprehensive search space of candidate learned sector classifications, we generated HCA models parameterized with each of the linkage methods, and then isolated sector classifications for varying numbers of sectors. Specifically, we varied the number of sectors in our search space universes with $N = \{5, 6, \ldots, 19 \}$ for each of the four linkage methods, for a total of 60 candidate learned sector universes. A subset of our search space is visualized in Figure~\ref{fig:hierarchical_clustering_model:partial_search_space}.

The HCA models were generated iteratively using the model input data discussed in Section~\ref{model_data}. We utilized the built-in Hierarchical Clustering Module in \textit{Scikit-learn}\citeFormat{\cite{Pedregosa2011Scikit-learn:Python}} to generate the models, and saved them in specially formatted CSV files (publicly available on the \textit{reIndexer} website\citeFormat{\cite{Weerawarana2019ReIndexerUniverses}}) for later ingestion by the ranking system we developed to identify the optimal learned sector classification universe.

% Candidate Universe Ranking

\chapter{Candidate Universe Ranking} \label{candidate_universe_ranking}
    
Following the successful derivation of an objective sector classification heuristic (addressing RG-1), the next step is to derive a methodology to rank our sectors against each other, to isolate the optimal sector/sectors, without imposing any subjective criteria on the selection. Thus, this section begins the discussion of addressing the second research goal, RG-2 (see Section~\ref{research_goals:specific_research_goals}):

\begin{table}[h!]
    \centering
    \begin{tabular}{| c | c |}
        \hline
        &  \\
        RG-2 & Rank candidate sector universes against each other using entirely objective criteria. \\
        & \\
        \hline
    \end{tabular}
\end{table}

\section{Sample Learned Sector Universe} \label{candidate_universe_ranking:sample_ls}

\begin{wrapfigure}[21]{l}{0.6\textwidth}
    \centering
    \vspace{\wrapfigadjustment}
    \fbox{
    \includegraphics[width=.9\linewidth]{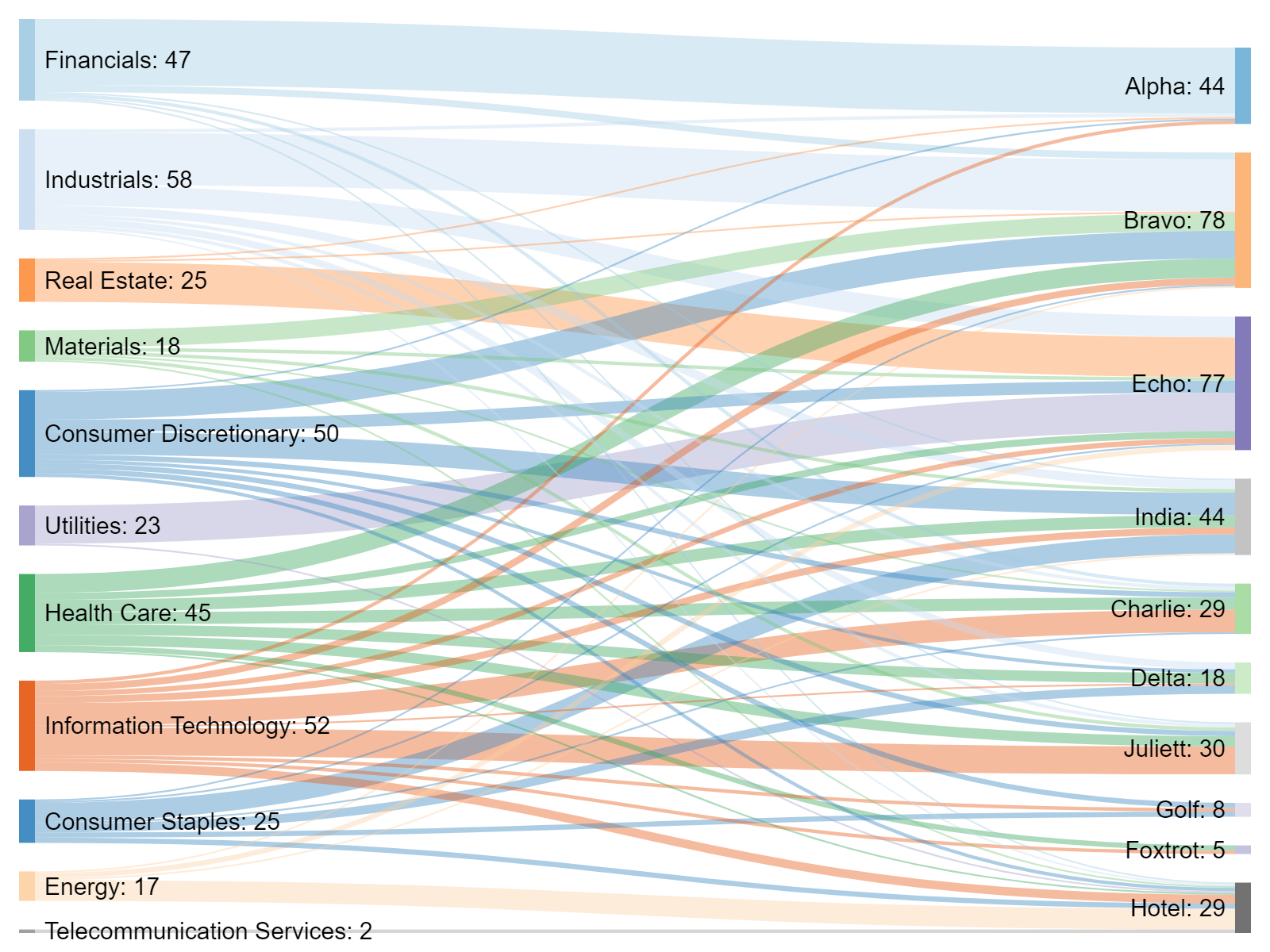}
    }
    \caption{Sample learned sector universe - Ward Linkage; 2010 Data; 10 Sectors.}
    \label{fig:candidate_universe_ranking:sample_ls_universe}
\end{wrapfigure}

To best motivate the approach we decided to use when comparing and ranking candidate sector universes, it is worthwhile to discuss an example.

Figure~\ref{fig:candidate_universe_ranking:sample_ls_universe} is a Sankey Diagram, representing the new learned sector assignments of various corporations from our search space by means of comparison to the benchmark. The left-hand side of the diagram represents the original benchmark sectors, and their constituent assets, while the right hand side represents the new learned sector assignment of the same assets.

As evidenced by the diagram, there appears to be significant transitory behavior of corporations across sectors when comparing the benchmark sectors to the learned sectors. Additionally, there also appears to be a large amount of mixing between the sectors, with very few sectors appearing to be preserved between the sector universes. This implies a stark lack of congruity between the benchmark classification and the learned sector universe heuristic. In the example, the only sector that can be considered remotely similar to a benchmark sector would be learned sector \textit{Alpha} to benchmark sector \textit{Financials}.

Other sectors however, appear largely broken up and dispersed when comparing their benchmark sector to the new learned sector universe. Particularly noticeable examples of this include the benchmark sectors \textit{Health Care}, \textit{Information Technology}, and \textit{Consumer Discretionary}. Due to the fundamentals-driven nature of our classification heuristic, this result is not entirely surprising; the benchmark sectors that appear to have the most dispersion in the learned sector universe are ones which are increasingly integral to regular business, regardless of sector; particularly \textit{Information Technology}.

As illustrated in this section with a single example, there is extremely little congruity between the benchmark sector classification and the learned sector classification. This pattern can be observed across a larger set of learned sector universes in Figure~\ref{fig:hierarchical_clustering_model:partial_search_space}, a partial visualization of the learned sector search space.

Due to this fact, is would be extremely difficult to perform a sector-by-sector analysis across sector universes as a means for comparison. There is obvious difficulty in matching sectors across universes (as illustrated by the example in Figure~\ref{fig:candidate_universe_ranking:sample_ls_universe}) without introducing significant bias to the comparison metric. Furthermore, there is an additional issue presented by the fact that the number of sectors in two given candidate learned sector universes may not be identical (let alone the number of constituent corporations in a given sector), thus completely prohibiting a sector-by-sector analysis.

To combat this issue, we decided to evaluate the sector universes as a whole, and then analyze universe-level metrics to rank the candidate learned sectors. To this end, we developed \textit{reIndexer}, which is discussed at length in the following section.

\section{reIndexer} \label{candidate_universe_ranking:reindexer}

\begin{wrapfigure}[18]{r}{0.4\textwidth}
    \centering
    \vspace{\wrapfigadjustment}
    \fbox{
    \includegraphics[width=.9\linewidth]{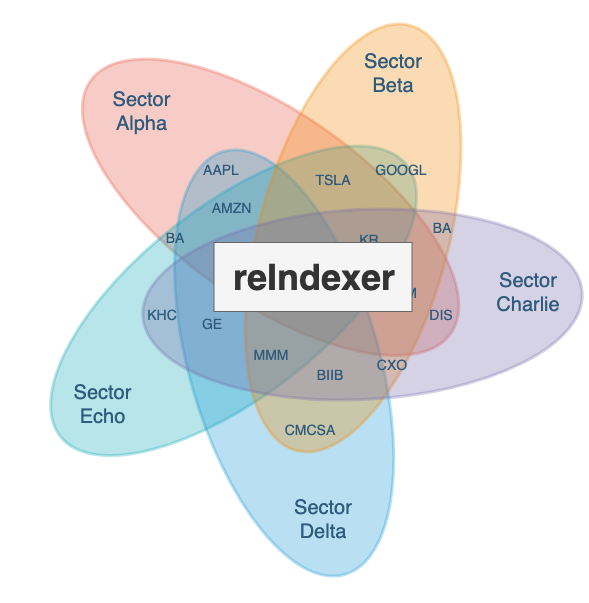}
    }
    \caption{The \textit{reIndexer} backtest-driven sector universe evaluation research tool.}
    \label{fig:candidate_universe_ranking:reindexer_logo}
\end{wrapfigure}

\textit{reIndexer}\citeFormat{\cite{Weerawarana2019ReIndexerUniverses}} is an open-source research tool for backtest-driven evaluation of different sector universes, using a system of Synthetic ETFs (hereafter \textit{SETF}), and efficient portfolios of those SETFs. reIndexer was designed and implemented to solve the problem discussed in the previous section; namely, the fact that we cannot perform a sector-by-sector comparison, and thus must compare learned sectors at the universe-level.

Built on Quantopian's \textit{Zipline Pythonic Algorithmic Trading Library}\citeFormat{\cite{QuantopianInc.2019ZiplineLibrary}}, reIndexer is fully API-compatible with the Quantopian suite of analytics tools, including \textit{Alphalens}\citeFormat{\cite{QuantopianInc.2019AlphalensFactors}} and \textit{Pyfolio}\citeFormat{\cite{QuantopianInc.2019PyfolioPython}}. While these libraries are not directly used in this project, the modular design of reIndexer make it an extremely powerful platform that is highly extensible in scope and functionality.

reIndexer was designed under the the hypothesis that the universe-level statistics of a given sectorization scheme would provide a superior metric for comparison across sector universes, compared to a sector-by-sector analysis. To this end - at a very high level - it provides a level of abstraction between single-asset trades, and constructed SETFs (whose composition is provided by the user), to simulate and record the performance of a portfolio of these SETFs over a predefined backtesting window.

\subsection{Synthetic ETF Formulation}

As the assets prescribed by the classification heuristic are not traded in the real market, it is not possible to get historical ETF prices directly from Zipline's engine. reIndexer works around this fact by implementing a layer between the portfolio optimization, and the assets, maintaining a hypothetical SETF. To compute efficient portfolios, and to treat the SETFs as unique assets from the point of view of the portfolio optimization engine, all that is necessary is a historical chain of prices for a given asset, over a specified lookback period.

To this end, reIndexer currently implements a price-weighted synthetic ETF. The pythonic implementation of this ETF is based on a highly flexible and reproducible tempalte, ensuring that different types of ETFs (market-weighted, etc.) can be easily implemented and used with reIndexer. For this project, price-weighted SETFs were used due to the fact that Zipline does not have historical market capitalization data for certain firms in its asset universe.

The mathematical formulation of the price-weighted SETF is outlined below:

\pagebreak

\begin{gather*}
    \text{Let $\boldsymbol{S}_\alpha$} = \text{Assets in sample sector $\alpha$} \\
    \text{Let $\boldsymbol{P}_\alpha$} = \text{Prices of assets in $\boldsymbol{S}_\alpha$} \\
    \text{Let $\boldsymbol{w}_\alpha$} = \text{Weights of assets in $\boldsymbol{S}_\alpha$} \\
    \\
    \Rightarrow \boldsymbol{w}_\alpha = \left[ \frac{P_i}{\sum \boldsymbol{P}_\alpha} \, \forall \, P_i \in \boldsymbol{P}_\alpha \right]^\intercal
\end{gather*}

This process of recomputing the weights of the constituent assets in a SETF is referred to as \textbf{SETF Restructuring}.

Suppose that this SETF restructuring process occurs at every time step $\{ r_t, r_{t+1}, \ldots, r_{t+n} \}$.

Additionally, let there be additional time steps in-between the restructuring process times, $\{ r_{t + \delta t}, r_{t + 2\delta t}, \ldots, r_{t + m \delta t} \}$, such that {${r_t < \{ r_{t + \delta t}, \ldots, r_{t + m\delta t} \} \leq r_{t + 1}}$}.

At each of these intermediate timesteps, the price-weighted SETF will have a price equal to the dot product of the sector asset weights computed at the immediately preceding SETF restructuring time $r_{t}$, $\boldsymbol{w}_{\alpha,r_t}$ and the prices of the constituent assets at the intermediate timestep, $\boldsymbol{P}_{\alpha,r_{t+m\delta t}}$.

\begin{gather*}
    \text{Let $\Pi_{\alpha,\tau}$} = \text{Price of sector SETF $\alpha$ at time $\tau$} \\
    \\
    \Rightarrow \Pi_{\alpha,\tau_i} = \boldsymbol{w}_{\alpha, r_t} \cdot \boldsymbol{P}_{\alpha,\tau_i} \, \forall \, \tau_i \in \{ r_{t + \delta t}, r_{t + 2\delta t}, \ldots, r_{t + m \delta t} \}
\end{gather*}

reIndexer provides an extremely flexible interface to specify SETF restructuring trigger times, $\{ r_t, r_{t+1}, \ldots, r_{t+n} \}$. The restructure can be triggered on any specific day of any specific week of the month (eg: third Friday of each month), or simply the first trading day of each month. Additionally, it also handles intelligent rebalancing rollovers, in the event that the specified rebalancing date trigger is a holiday with respect to the configured trading calendar.

\subsection{Efficient Portfolio Optimization} \label{candidate_universe_ranking:port_optim}

Following the construction of the SETFs, reIndexer builds an efficient portfolio, treating each of the SETFs as distinct, unitary assets. As our goal with the project is to assess the benefit of fundamentals-driven objective sector classifications through the lens of risk diversification, reIndexer currently implements a backtest of a Global Minimum Variance Portfolio, not allowing short-sales.

However, in a similar fashion to the price-weighted SETF, the pythonic implementation of this portfolio is highly generalized in the reIndexer source code, and can be easily reconfigurable to work with myriad different portfolio configurations. reIndexer provides functionality to retrieve a Matrix of historical SETF prices for a given lookback window, with SETF prices being correctly computed using the formulation described above.

Following this, reIndexer computes the correlation matrix of returns using the historical prices over a specific lookback period, and then performs the non-convex optimization necessary to compute SETF weights in the Global Minimum Variance Portfolio. A sequential quadratic programming solver from the Python library \textit{SciPy}\citeFormat{\cite{Oliphant2007PythonComputing}} was used to perform the optimization.

The mathematical formulation for this process is outlined below (note that notation is preserved from the preceding section):

\pagebreak

\begin{gather*}
    \text{Let $\Omega$} = \text{Set of sectors in the candidate universe, $\Omega_i \in \{\Omega_1, \Omega_2, \ldots, \Omega_n\}$} \\
    \text{Let $\boldsymbol{\Pi}_{\Omega_i}$} = \text{Historical log price vector of sector SETF $\Omega_i$} \\
    \text{Let $\boldsymbol{\Sigma}$} = \text{Covariance matrix of historical log-returns of SETFs } \\
    \text{Let $\boldsymbol{\omega}$} = \text{Vector of SETF weights in the Global Minimum Variance Portfolio} \\
    \\
    \boldsymbol{\Sigma} =
    \begin{bmatrix}
        \mathbb{E}[(\boldsymbol{\Pi}_{\Omega_1} - \mathbb{E}[\boldsymbol{\Pi}_{\Omega_1}])^2] & \mathbb{E}[(\boldsymbol{\Pi}_{\Omega_1} - \mathbb{E}[\boldsymbol{\Pi}_{\Omega_1}])(\boldsymbol{\Pi}_{\Omega_2} - \mathbb{E}[\boldsymbol{\Pi}_{\Omega_2}])] & \cdots & \mathbb{E}[(\boldsymbol{\Pi}_{\Omega_1} - \mathbb{E}[\boldsymbol{\Pi}_{\Omega_1}])(\boldsymbol{\Pi}_{\Omega_n} - \mathbb{E}[\boldsymbol{\Pi}_{\Omega_n}])] \\
        & & & \\
        \mathbb{E}[(\boldsymbol{\Pi}_{\Omega_2} - \mathbb{E}[\boldsymbol{\Pi}_{\Omega_2}])(\boldsymbol{\Pi}_{\Omega_1} - \mathbb{E}[\boldsymbol{\Pi}_{\Omega_1}])] & \mathbb{E}[(\boldsymbol{\Pi}_{\Omega_2} - \mathbb{E}[\boldsymbol{\Pi}_{\Omega_2}])^2] & \cdots & \mathbb{E}[(\boldsymbol{\Pi}_{\Omega_2} - \mathbb{E}[\boldsymbol{\Pi}_{\Omega_2}])(\boldsymbol{\Pi}_{\Omega_n} - \mathbb{E}[\boldsymbol{\Pi}_{\Omega_n}])] \\
        & & & \\
        \vdots & \vdots & \ddots & \vdots \\
        & & & \\
        \mathbb{E}[(\boldsymbol{\Pi}_{\Omega_n} - \mathbb{E}[\boldsymbol{\Pi}_{\Omega_n}])(\boldsymbol{\Pi}_{\Omega_1} - \mathbb{E}[\boldsymbol{\Pi}_{\Omega_1}])] & \mathbb{E}[(\boldsymbol{\Pi}_{\Omega_n} - \mathbb{E}[\boldsymbol{\Pi}_{\Omega_n}])(\boldsymbol{\Pi}_{\Omega_2} - \mathbb{E}[\boldsymbol{\Pi}_{\Omega_2}])] & \cdots & \mathbb{E}[(\boldsymbol{\Pi}_{\Omega_n} - \mathbb{E}[\boldsymbol{\Pi}_{\Omega_n}])^2] \\
    \end{bmatrix}
    \\
    \\
    \therefore \, \text{Global minimum variance portfolio weights are determined by solving the non-convex optimization:}
\end{gather*}

\begin{equation*}
    \begin{aligned}
        & \underset{\boldsymbol{\omega}}{\text{minimize}} & & \boldsymbol{\omega}^\intercal, \boldsymbol{\Sigma} \, \boldsymbol{\omega} \\
        & \text{subject to}
        & & \boldsymbol{1}^\intercal \cdot \boldsymbol{\omega} = 1 \\
        & & & \omega_i \geq 0 \, \forall \,\omega_i \in \boldsymbol{\omega}
    \end{aligned}
\end{equation*}

This process of recomputing the weights of the SETFs in the portfolio is referred to as \textbf{Portfolio Rebalancing}.

Similar to the SETF restructuring process, the portfolio rebalanacing process too occurs at discrete, user-specified time intervals. Additionally, as with the restructuring process, portfolio weights from the preceding timestep are used to compute the value of the portfolio at each intermediate timestep. This computation is outlined below:

\begin{gather*}
    \text{Let $\boldsymbol{\omega_{\tau}}$} = \text{SETF portfolio weights at time $\tau$} \\
    \text{Let $\pi_{\boldsymbol{\Omega},\tau}$} = \text{Value of portfolio under sector universe $\boldsymbol{\Omega}$ at time $\tau$} \\
    \\
    \\
    \Rightarrow \forall \, \{ i, \, j, \, \tau \} \, : \{ \{ i \in \boldsymbol{\Omega} \}; \, \{ j \in \boldsymbol{S_i} \}; \, \{ r_{t} < \tau \leq r_{t+m\delta t} < r_{t + 1} \} \}
\end{gather*}
\begin{align*}
    \therefore \pi_{\boldsymbol{\Omega}, \tau} 
    =& \; \Pi_{\boldsymbol{\Omega}} \cdot \boldsymbol{\omega} \\
    =& \sum_{\boldsymbol{i \in \Omega}} \Pi_{\Omega_i, \tau} \cdot \omega_{i, \tau} \\
    =& \sum_{i \in \boldsymbol{\Omega}} \boldsymbol{w}_{i, \tau} \cdot \boldsymbol{P}_{i, \tau} \cdot \omega_{i, \tau} \\
    =& \sum_{i \in \boldsymbol{\Omega}} \sum_{j \in \boldsymbol{S}_i} w_{j, \tau} \cdot P_{j, \tau} \cdot \omega_{i, \tau}
\end{align*}

In addition to the portfolio value, other metrics are also recorded at each timestep. These include open positions, the Sharpe ratio, the information ratio, and myriad other portfolio statistics. In this project, we do not utilize the full battery of statistics provided by Zipline (over 30 individual statistics in total), but we maintain the capability to generate these statistics for future expansion of reIndexer, and to maintain API compatibility with the Quantopian suite of analysis tools.

\pagebreak

\subsection{Software Architecture Overview}

In this section, we describe the software architecture of reIndexer. Figure~\ref{fig:candidate_universe_ranking:reindexer_architecture} is the architecture diagram of the system, describing the logical flow of the system, from individual asset statistics on the left, to final trades executed with the Zipline engine on the right.

\begin{figure}[h!]
    \centering
    \fbox{
    \includegraphics[width=.8\linewidth]{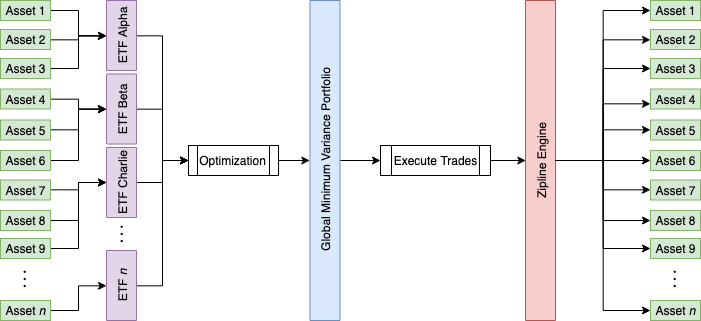}
    }
    \caption{reIndexer architecture overview diagram.}
    \label{fig:candidate_universe_ranking:reindexer_architecture}
\end{figure}

As seen in the diagram, SETFs are constructed from the bare asset statistics on the left, and are treated as unitary assets. Upon computation of SETF historical price chains (outlined above), the Global Minimum Variance Portfolio (with no short-sales) is computed by solving the non-convex optimization, also outlined above.

Both of the intervals of computation for these distinct determinants of portfolio value, the asset weights in the SETF, and the SETF weights in the portfolio, are computed at discrete, user-configurable intervals and do not necessarily have to occur at the same time. This behavior allows to replicate the real market to a high degree of accuracy, as most popular ETFs are rebalanced on the third Friday of every month, whereas retail investor portfolios are typically rebalanced on the first trading day of each month. Furthermore, Zipline also maintains the full historical order book for each asset, thus providing realistic price drift effects when large orders are placed.

After both computations are performed, individual trades are passed to the Zipline layer of the system (highlighted in red). These trades are only executed on the days of either SETF restructuring (i.e. $\boldsymbol{w}$ update) or portfolio rebalancing (i.e. $\boldsymbol{\omega}$ update). This provides us with a significant performance boost, as Zipline is optimized to record historical portfolio performance extremely quickly when trade execution is not required.

As Zipline does not see the individual SETFs (they are maintained internally by reIndexer), it is necessary to compute the individual weight for a given asset when a trade is to be executed. Due to the fact that the portfolio of SETFs can be considered a portfolio of portfolios, which both have asset weights that sum to 1, the weight of an individual asset in the larger portfolio of SETFs is simply the product of its weight in the SETF, and the weight of the SETF in the global minimum variance portfolio. This computation is outlined below:

\begin{gather*}
    \text{Let $\boldsymbol{\Theta}$} = \text{Set of all assets in the simulation} \\
    \text{Let $\boldsymbol{w}_i(\theta)$} = \text{Weight of asset $\theta$ in SETF $i$} \\
    \text{Let $\boldsymbol{\omega}(\theta, i)$} = \text{Weight of SETF $i$ in the Global Minimum Variance Portfolio} \\
    \text{Let $\gamma_\theta$} = \text{True weight of asset $\theta$ in the larger Zipline portfolio}
    \\
    \\
    \therefore \, \gamma_\theta = \boldsymbol{w}_i(\theta) \cdot \boldsymbol{\omega}(\theta, i) \; \forall \; \{ \theta, i \} : \{ \{ \theta \in \boldsymbol{\Theta} \} ; \, \{ i \in \boldsymbol{\Omega} \} \}
\end{gather*}

\section{Performance Evaluation Metrics} \label{candidate_universe_ranking:eval_metrics}

To fully address RG-2 (see Section~\ref{research_goals:specific_research_goals}), we selected a set of objective criteria to record over the duration of each backtest, for each candidate learned sector universe. These metrics were selected to evaluate specific, but varied attributes of learned sector universe SETF portfolios, and will be evaluated independently to identify the optimal learned sector universe with respect to each metric.

\subsection{SETF Restructuring Turnover}

As described above, a SETF event is the recomputation of constituent asset weights in a synthetic ETF, and is a user-configurable triggered event. In the financial markets today, sector ETFs are typically created and sold by financial institutions. This abundance of ETFs provides increased liquidity to the market, while also reducing the barrier for entry to retail investors to gain exposure to specific sectors.

A key cost of creating and holding these ETFs for financial institutions is the fee incurred during ETF restructuring. Due to their enhanced status in the market, large financial institutions do not pay traditional commission fees charged to retail investors when executing trades on a stock exchange.

However, we believe it is valid to assume that their cost of trading would be proportional to the asset turnover of the component assets at each of these restructuring times. Therefore, we chose to record the component SETF asset restructuring turnover at each time a restructure is triggered, and utilize it as a proxy for judging the cost to financial institutions that would be incentivized to create ETFs of these sectors if they were indeed real. The mathematical formulation of this turnover for a single restructuring event is outlined below:

\begin{gather*}
    \text{Let $r_t$ and $r_{t+1}$} = \text{Times SETF restructures are triggered} \\
    \text{Let $\boldsymbol{w}_\tau$} = \text{Vector of underlying asset weights in the SETF at time $\tau$} \\
    \text{Let $\boldsymbol{P}_\tau$} = \text{Vector of underlying SETF asset prices at time $\tau$} \\
    \\
    \Rightarrow \text{SETF Restructuring Turnover} = \left| \boldsymbol{w}_{r_{t+1}}^\intercal - \boldsymbol{w}_{r_{t}}^\intercal \right| \cdot \boldsymbol{P}_{r_{t + 1}}
\end{gather*}

\subsection{Portfolio Rebalancing Turnover}

Similar to how we treat the SETF restructuring turnover as a proxy for the cost to a financial institution to create the SETFs in the real market, we treat the portfolio rebalancing turnover as a proxy for the cost to a retail investor holding the SETFs in a portfolio.

To avoid introducing specific cost bias to our evaluation, we simply record the turnover of each SETF at each portfolio rebalance event. The mathematical formulation of this turnover for a single rebalancing event is outlined below:

\begin{gather*}
    \text{Let $\tau_t$ and $\tau_{t+1}$} = \text{Times portfolio rebalancing events are triggered} \\
    \text{Let $\boldsymbol{\omega}_t$} = \text{Vector of SETF weights in the portfolio at time $t$} \\
    \text{Let $\boldsymbol{\pi}_t$} = \text{Vector of prices of SETFs in the portfolio at time $t$} \\
    \\
    \Rightarrow \text{Portfolio Restructuring Turnover} = \left| \boldsymbol{\omega}_{\tau_{t + 1}}^\intercal - \boldsymbol{\omega}_{\tau_t}^\intercal \right| \cdot \boldsymbol{\pi}_{\tau_{t+1}}
\end{gather*}

\pagebreak

\subsection{Portfolio Return}

We also record the overall portfolio return over time. The computation of portfolio value at each timestep is discussed in depth in the previous section.

\subsection{Sharpe Ratio}

To best evaluate our initial hypothesis that our learned sector classification heuristic - based on objective fundamentals data - would provide better economically diversified sector classifications, we must test the diversification benefit of our learned sector universes, relative to the benchmark. Similarly, we also utilize this metric to isolate the optimal learned sector universe portfolio with respect to this diversification benefit metric.

As we are already computing the global minimum variance portfolio with no short sales, the total variance of the portfolio is minimized. Due to this fact, we can treat the risk-adjusted return (i.e. the Sharpe Ratio) as a metric for quantifying the diversification benefit of each learned sector classification universe with respect to a single unit of risk. The mathematical formulation of risk-adjusted return captured by reIndexer is reproduced below (parameterized by the annualized portfolio return, $R_p$ and the portfolio variance $\sigma_p$):

\begin{gather*}
    \text{Sharpe Ratio} = \frac{R_p - r_f}{\sigma_p}
\end{gather*}

\section{Backtest Configuration} \label{candidate_universe_ranking:backtest_config}

A backtest was performed on all 60 candidate learned sector universes with the following configuration:

\begin{table}[h!]
    \centering
    \begin{tabular}{|c|c|}
        \hline
        \textbf{Configuration Parameter} & \textbf{Setting} \\
        \hline
        Start Date & January 1, 2012 \\
        End Date & December 31, 2017 \\
        SETF Restructure Trigger & Third Friday of each month \\
        Portfolio Rebalance Trigger & First trading day of each month \\
        Starting Capital & \$10,000,000,000 \\
        Backtest Frequency & Daily \\
        \hline
    \end{tabular}
    \caption{Learned sector universe backtest configuration parameters.}
    \label{table:candidate_universe_ranking:backtest_configuration}
\end{table}

\textit{Note: A high initial capital base is necessary to account for the fact that Zipline does not allow for fractional asset trades. Using a large capital base mitigates the rounding effect caused by this behavior.}

% Optimal Sector Universes

\chapter{Optimal Sector Universes} \label{optimal_sector_universes}

In this chapter, we utilize the backtesting system outlined in the previous chapter, we performed historical universe-level analysis of each of our candidate learned sectors. To reiterate, we plan to use the SETF restructuring turnover, the portfolio rebalancing turnover, the portfolio value, and the Sharpe ratio to rank our candidate learned sectors against each other.

Due to the magnitude of data output by the reIndexer backtesting system, all calculations in this section were performed in the cloud, on Google Research Colaboratory.\citeFormat{\cite{GoogleResearch2019GoogleColaboratory}} Data was loaded dynamically from its output location on Google Drive, directly into Google Colaboratory. Following this, individual output files were opened sequentially to extract the necessary longitudinal data dimension for the candidate learned sector universe, with all data being kept in-memory on Google Cloud Platform servers for increased efficiency during analysis.\citeFormat{\cite{Weerawarana2019LearnedAnalytics}}

\section{Complete Backtest Results}

The complete results of the backtest are published online, and are available on the reIndexer website.\citeFormat{\cite{Weerawarana2019ReIndexerUniverses}}

Due to the magnitude of data generated by reIndexer, it is impractical to display quantitative summaries for each sector. Rather, we plotted graphs to visualize the progression of each of the risk metrics discussed in Section~\ref{candidate_universe_ranking:eval_metrics}. The graphs were plotted for the full backtesting window outlined in Section~\ref{candidate_universe_ranking:backtest_config}.

\begin{itemize}
    \item Cumulative SETF Restructuring Turnover (Figure~\ref{fig:appendix_backtest:setf_restructuring_turnover})
    \item Cumulative Portfolio Rebalancing Turnover (Figure~\ref{fig:appendix_backtest:portfolio_rebal_turnover})
    \item Portfolio Value (Figure~\ref{fig:appendix_backtest:portfolio_return})
    \item Rolling Sharpe Ratio (Figure~\ref{fig:appendix_backtest:sharpe_ratio})
\end{itemize}

See Appendix~\ref{appendix:backtest_visualization} on Page~\pageref{appendix:backtest_visualization} for full-sized plots of each of the graphs.

% \pagebreak

\section{Backtest Results Analysis}

We computed and recorded each of the Performance Evaluation Metrics outlined in Section~\ref{candidate_universe_ranking:eval_metrics} for each time step across all 60 candidate learned sector universes.

Following this, we computationally extract the optimally performing learned sector universe for each of the performance metrics. As both turnover metrics track a cost, we isolated the learned sector universe with the minimum cumulative value at the end of the simulation. Conversely, we isolated the learned sector universe that yielded the maximum value for both the portfolio value, and the (average) rolling Sharpe ratio.

We now discuss the optimally performing learned sector universe with respect to the turnover and maximum absolute portfolio value metrics. Coincidentally, the minimum cumulative SETF restructuring turnover, and the minimum portfolio rebalancing turnover are both achieved by the same learned sector universe. These two performance metrics will be discussed together.

\subsection{Minimum Cumulative Turnover}

\begin{wrapfigure}[22]{r}{0.5\textwidth}
    \centering
    \vspace{\wrapfigadjustment}
    \fbox{
    \includegraphics[width=.9\linewidth]{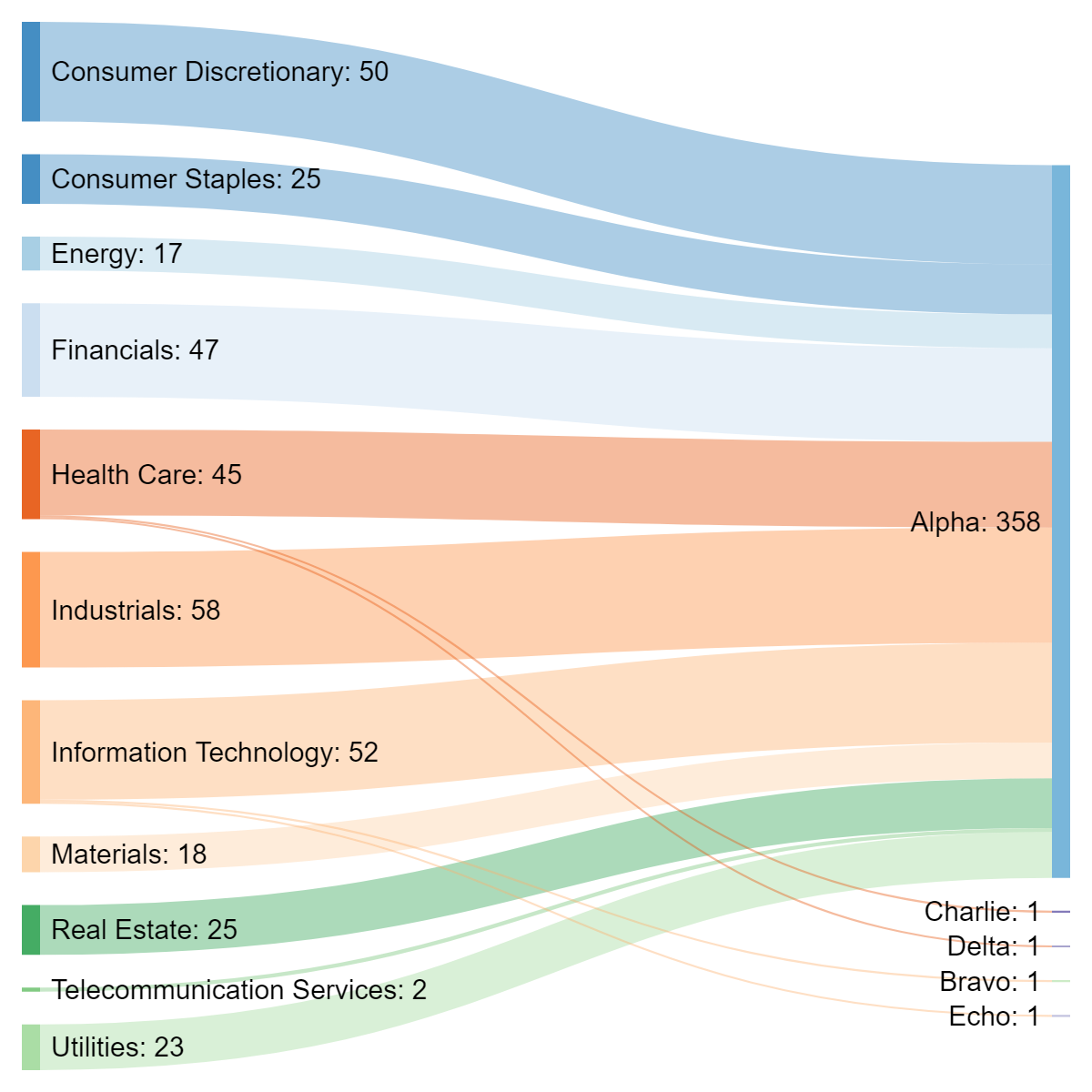}
    }
    \caption{Learned sector universe with minimum cumulative SETF restructuring, and portfolio reblancing turnover.}
    \label{fig:optimal_sector_universe:minimum_turnover}
\end{wrapfigure}

As discussed in Section~\ref{candidate_universe_ranking:eval_metrics}, the cumulative turnover, for both SETF restructuring, and portfolio rebalancing, are designed to be a proxy for the cost of issuing, and holding these ETFs, respectively. The learned sector that performed best with respect to the cumulative turnover metrics has the following configuration:

\vspace{0.5em}

\begin{minipage}{\linewidth}
    \centering
    \bfseries
    \fbox{
        Single Linkage; 5 Sectors
    }
\end{minipage}

\vspace{0.5em}

The cost proxies each measure the effective cost of the SETF for entirely different segments of the Financial Lifecycle of this hypothetical product. The SETF restructuring turnover is a proxy for the cost that would be incurred by an ETF issuer (i.e. an institutional investor), whereas the portfolio rebalancing turnover is a proxy for the cost incurred by a holder of the ETF (i.e. a retail investor).

Upon closer inspection of the Sankey diagram corresponding to this sector in Figure~\ref{fig:optimal_sector_universe:minimum_turnover}, it is clear why it has superior minimum turnover, through the lens of both SETF restructuring, and portfolio rebalancing. The presence of a single large sector containing most of the assets (all of the assets with the exception of 4 in the case of Figure~\ref{fig:optimal_sector_universe:minimum_turnover}) implies that the sector universe would inherently have a significant advantage over its counterparts.

This single-sector pooling behavior would imply its SETF restructuring fee would be 0 in perpetuity for the sectors in which there is only a single asset. As single-asset sectors represent 80\% of the sectors in this universe, this result is not surprising. This advantage is conferred in a similar fashion to the portfolio restructuring fee, as single-asset sector SETFs are not likely to be held in large quantities, relative to the singular vast sector SETF (Sector \textit{Alpha}, in this example).

Unfortunately, it seems that the single linkage method results in a significant \textit{pooling} effect, where the vast majority of assets are relegated to a single sector, with the remaining sectors being relative extremely small; containing just one asset each in the case of Figure~\ref{fig:optimal_sector_universe:minimum_turnover}. This behavior under the single linkage method is apparent from both the underlying data, as well as the partial search space visualization in Figure~\ref{fig:hierarchical_clustering_model:partial_search_space}.

% \pagebreak

\subsection{Maximum Absolute Portfolio Value}

\begin{wrapfigure}[23]{L}{0.5\textwidth}
    \centering
    \fbox{
    \includegraphics[width=.9\linewidth]{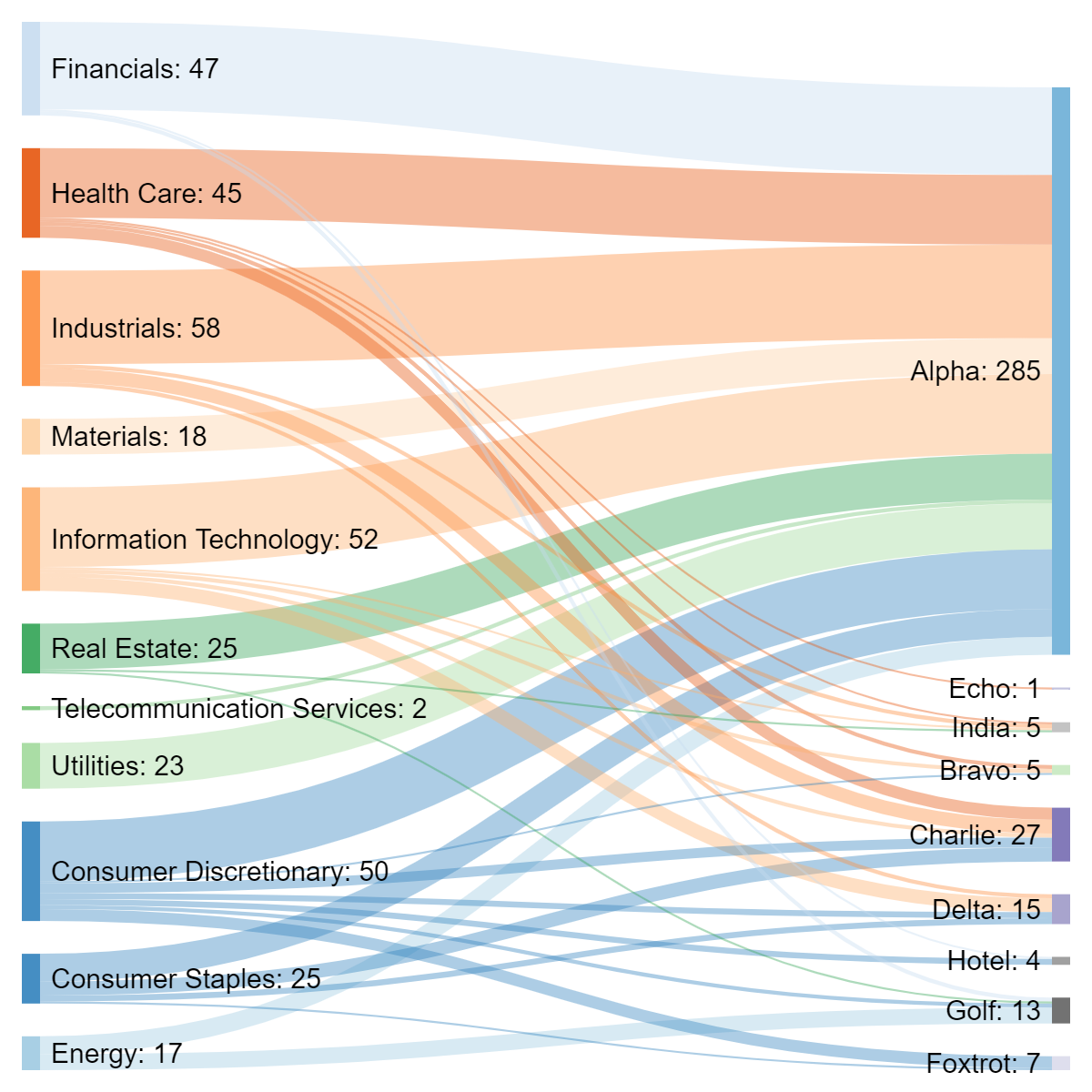}
    }
    \caption{Learned sector universe with maximum absolute portfolio value.}
    \label{fig:optimal_sector_universe:max_return}
\end{wrapfigure}

Unlike the previous performance metric, evaluating the historical learned sector universe portfolio value is not a proxy for an external cost. Rather, it is a candid assessment of the historical performance of the learned sector universes. While this does not directly address any of our research goals, it is an extremely important performance metric, and would be a key determinant of the success of any given learned sector universe in the real market. The learned sector that had the maximum absolute portfolio value over the lookback period (outlined in Section~\ref{candidate_universe_ranking:backtest_config}) has the following configuration:

\vspace{.5em}

\begin{minipage}{\linewidth}
    \centering
    \bfseries
    \fbox{
        Complete Linkage; 9 Sectors
    }
\end{minipage}

\vspace{.5em}

The absolute portfolio value tracks the value of the portfolio throughout the backtesting period, utilizing the nested value computation outlined in Section~\ref{candidate_universe_ranking:port_optim}. While not being a direct proxy of risk diversification, if viewed through the lens of potentially improved economic asset grouping, this learned sector universe is extremely encouraging. If the corollary that better capital market performance is correlated with improved economic sector performance, this result would imply that the fundamentals-driven classification heuristic does indeed provide a beneficial measure of diversification.

Unlike the previous learned sector universe under the Single Linkage Method, the learned sector outlined in Figure~\ref{fig:optimal_sector_universe:max_return} appears to have significantly less pooling. Despite this however, assets are still highly concentrated in a single extremely large sector, with other sector sizes (with respect to the number of constituent companies) being significantly smaller.

An interesting observation of this sector is that there is significant dispersion from the original benchmark classification. Omitting the sector \textit{Alpha} due to its gargantuan size, sectors \textit{Charlie} and \textit{Delta} both have numerous components from highly diverse original benchmark sectors. Sector \textit{Charlie} appears to have a high number of assets from the \textit{Health Care} sector, as well as the \textit{Industrials} and \textit{Consumer Discretionary} sectors. Similarly, sector \textit{Delta} has a large number of constituent assets classified as \textit{Information Technology}, \textit{Consumer Staples}, and \textit{Consumer Discretionary} under the benchmark classification.

As with the sample sector classification discussed in Section~\ref{candidate_universe_ranking:sample_ls}, there appears to be high levels of dispersion of assets in sectors which are increasingly requisite to doing business. This is particularly evident (again, ignoring sector \textit{Alpha}) through the dispersion of the \textit{Information Technology}, \textit{Health Care}, and \textit{Consumer Discretionary} sectors.

\section{Risk-Adjusted Return Optimal Universe} \label{optimal_sector_universe:risk_adj_return_optimal}

Finally, to isolate the sector with the maximum rolling annualized Sharpe ratio, we computed the mean rolling Sharpe ratio across the longitudinal temporal axis, and compared each of the learned sectors. Following this comparison, we determined that the learned sector that had the maximum average rolling Sharpe Ratio over the lookback period has the following configuration:

\begin{minipage}{\linewidth}
    \centering
    \bfseries
    \fbox{
        Complete Linkage; 17 Sectors
    }
\end{minipage}

\vspace{1em}

The Sankey diagram corresponding to the risk-adjusted return optimal learned sector is displayed in Figure~\ref{fig:optimal_sector_universe:max_sharpe}. As indicated by the diagram, this learned sector universe has even less pooling behavior that both of the previously discussed learned sector universes (Figure~\ref{fig:optimal_sector_universe:minimum_turnover} and Figure~\ref{fig:optimal_sector_universe:max_return}), despite still having 2 relatively large sectors, \textit{Alpha}, and \textit{Golf}, when compared to the rest.

As this learned sector universe provides the best risk-adjusted return on an already variance-minimized Global Minimum Variance Portfolio comparison, it would imply that this configuration provides the best risk diversification profile out of all of the candidate learned sector universes. In addition to the obviously highly prevalent dispersion from the benchmark classification to the new learned sectors, the size of each of the sectors is also considerably more even compared to its peers. This would imply a better portfolio-level diversification profile, as poorly performing sectors can be underweight during times of low implied risk-adjusted return. This is a key caveat of the pooling behavior particularly prevalent under the Single Linkage heuristic.

This lack of transitivity, combined with the implication that a higher risk-adjusted return implies a better risk-diversification profile would suggest that the learned sector universe heuristic is providing high levels of economic diversification, which appears to be at odds with the benchmark classification.

The full dataset of the risk-adjusted return optimal learned sector universe is reproduced in Appendix~\ref{appendix:optimal_ls}.

\pagebreak

\begin{figure}
    \centering
    \fbox{
        \includegraphics[width=0.9\linewidth]{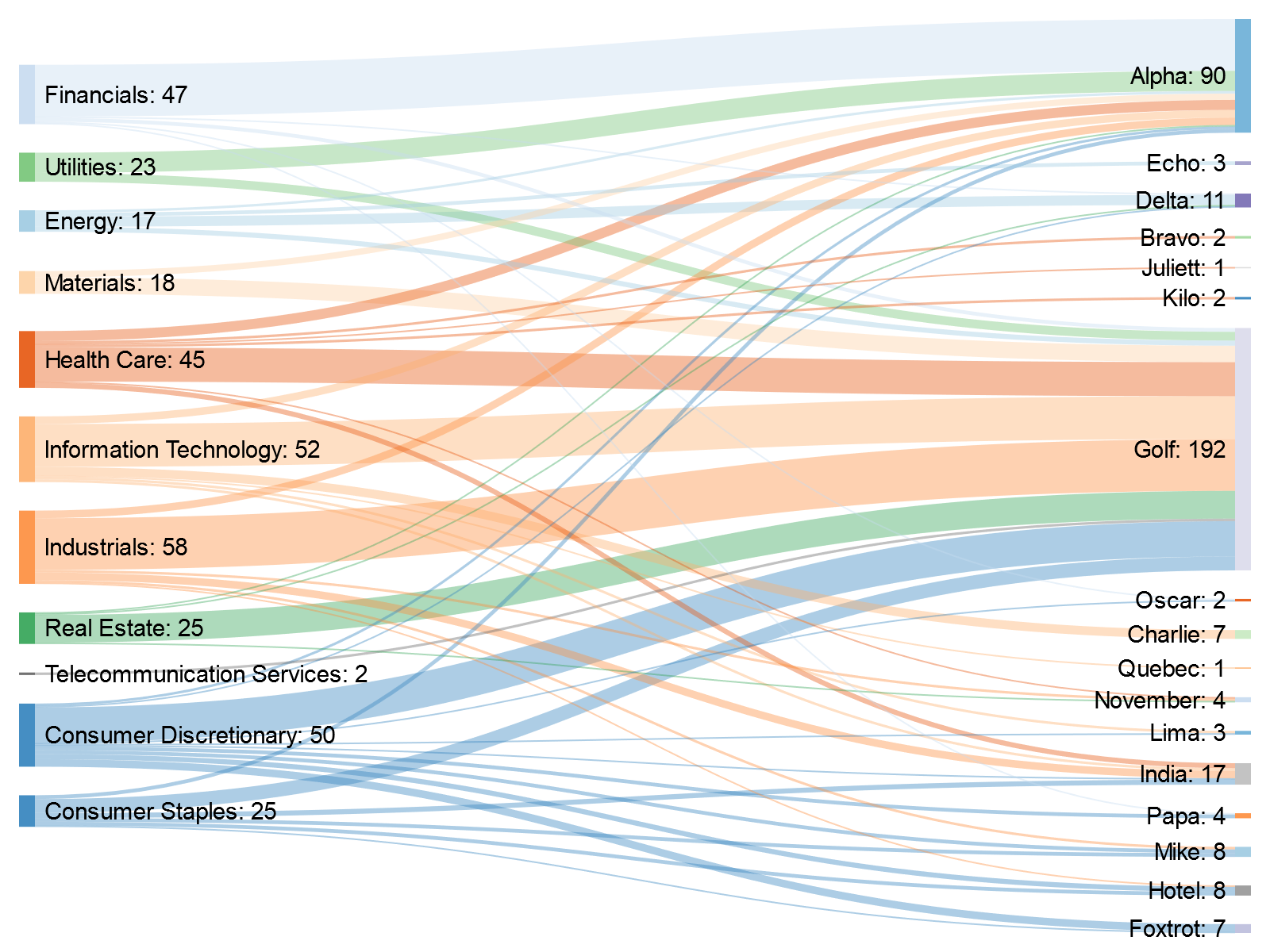}
    }
    \caption{Learned sector universe with maximum rolling Sharpe ratio.}
    \label{fig:optimal_sector_universe:max_sharpe}
\end{figure}

% Benchmark Comparison

\chapter{Benchmark Comparison} \label{benchmark_comparison}

We now have a data-driven methodology for deriving learned sector universes (addressing RG-1), and have developed a objective criteria-driven ranking methodology to compare the learned sector universes against each other, addressing RG-2. The final step is to evaluate our objectively-identified risk-adjusted return optimal learned sector universe (see Section~\ref{optimal_sector_universe:risk_adj_return_optimal}) against the benchmark classification. Thus, this section addresses the third and final research goal, RG-3 (see Section~\ref{research_goals:specific_research_goals}).

\begin{table}[h!]
    \centering
    \begin{tabular}{| c | c |}
        \hline
        &  \\
        RG-3 & Evaluate our risk-adjusted return optimal sector universe against the benchmark. \\
        & \\
        \hline
    \end{tabular}
\end{table}

\section{Comparison Overview}

To preserve the impartial basis for comparison developed and maintained throughout this report, we isolated sector assignments for our benchmark sector universe, the \textit{GICS S\&P 500 Classification}. Unfortunately, as mentioned previously, we were only able to isolate transverse sector assignments for the year 2019, and were unable to access historical sector assignments, thus making a truly longitudinal comparison of our learned sector to the benchmark impossible.

To mitigate this issue, we compared the latest learned sector as implied by our clustering algorithm to the benchmark classification. To maintain the consistency of our analysis, we utilized reIndexer (see Section~\ref{candidate_universe_ranking:reindexer}) to model SETFs of the benchmark, and to perform a backtest using the same configuration as was used for the candidate learned sector ranking (see Section~\ref{candidate_universe_ranking:backtest_config}). Similarly, we utilized the same performance metrics as were used to compare the candidate learned sector universes (see Section~\ref{candidate_universe_ranking:eval_metrics}) to compare the risk-adjusted return optimal learned sector universe to the benchmark.

\section{Performance Metric Comparison}

Figure~\ref{fig:benchmark_comparison:performance_metrics} contains four panels, (a) through (d), with each displaying one of the four performance metrics outlined in Section~\ref{candidate_universe_ranking:eval_metrics}. To best decompose the results of the comparison with the \textit{GICS S\&P 500 Classification} benchmark, we will analyze each of the performance metric comparisons in turn.

\subsection{Cumulative Turnover Comparison}

Panels (a) and (b) in Figure~\ref{fig:benchmark_comparison:performance_metrics} plot the cumulative turnover of SETF restructuring, and portfolio rebalancing, respectively. Recall from the previous section that for these turnover metrics, lower is better. As the red line represents the benchmark, it is apparent that the risk-adjusted return optimal learned sector did not outperform the benchmark with respect to both the SETF restructuring turnover, and the portfolio rebalancing turnover.

\pagebreak

This phenomenon may be explained through analysis of the Sankey Diagram of the risk-adjusted return learned sector universe in Figure~\ref{fig:optimal_sector_universe:max_sharpe}. It indicates a significant proliferation of component assets in 2 large sectors, with a large number of smaller sectors. Due to the fact that larger sectors have a higher notional value, and thus imply higher turnover when bought or sold, it is not surprising that the portfolio rebalancing turnover is higher for the risk-adjusted return optimal learned sector universe, compared to the more uniformly distributed benchmark universe.

Additionally, the larger individual sectors \textit{Alpha} and \textit{Golf} would also imply a higher rate of turnover during SETF restructuring. As a higher number of assets implies a more volatile total value, the extremely large sectors unique to the risk-adjusted return optimal learned sector universe would command a higher level of turnover during SETF restructuring, when compared to the more modestly sized sectors of the benchmark universe.

% See: https://tex.stackexchange.com/questions/238923/refer-to-a-table-as-figure
\begin{table}[!h]
    \centering
    % \fbox{
    \begin{tabular}{|c|c|}
        \hline
        & \\
        \includegraphics[width=.475\linewidth]{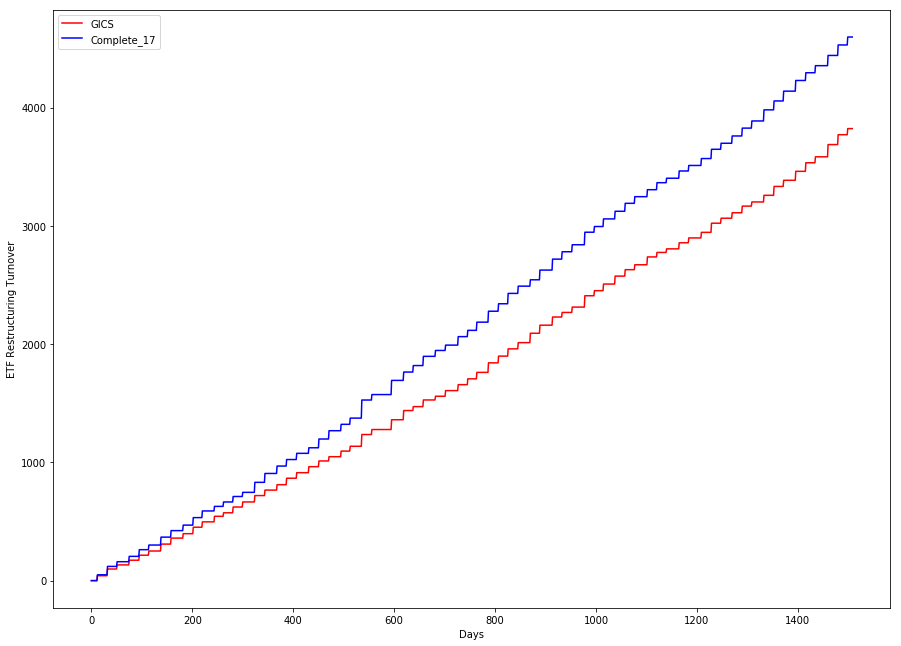} &
        \includegraphics[width=.475\linewidth]{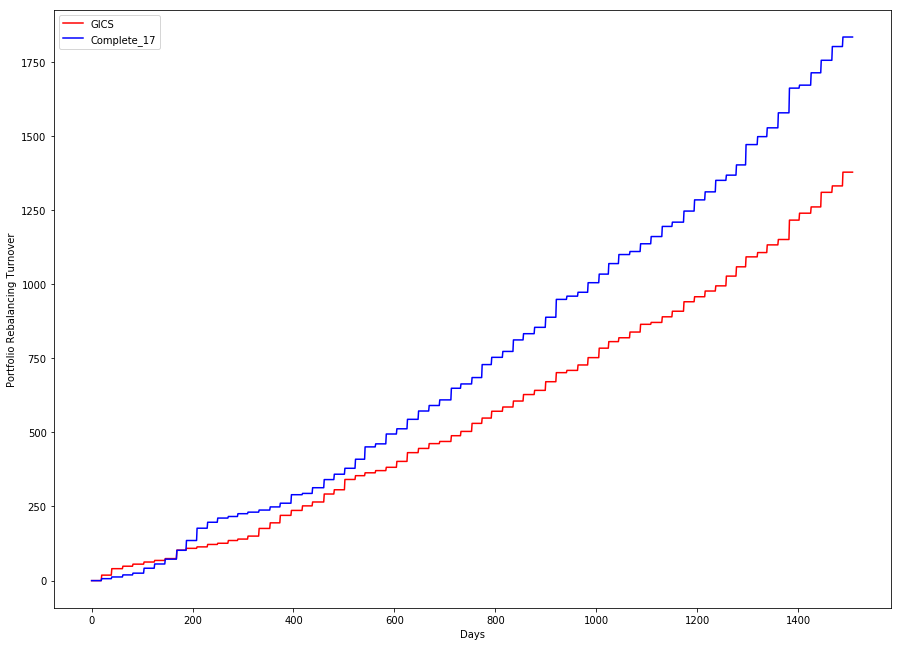} \\
        \textit{(a) SETF Restructuring Turnover} & \textit{(b) Portfolio Rebalancing Turnover} \\
        & \\
        \hline
        & \\
        \includegraphics[width=.475\linewidth]{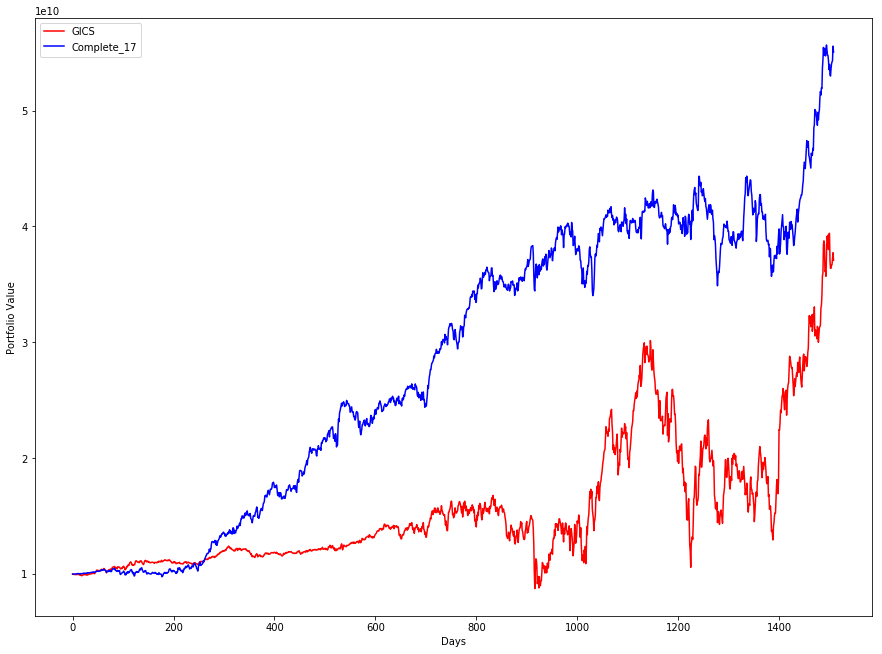} &
        \includegraphics[width=.475\linewidth]{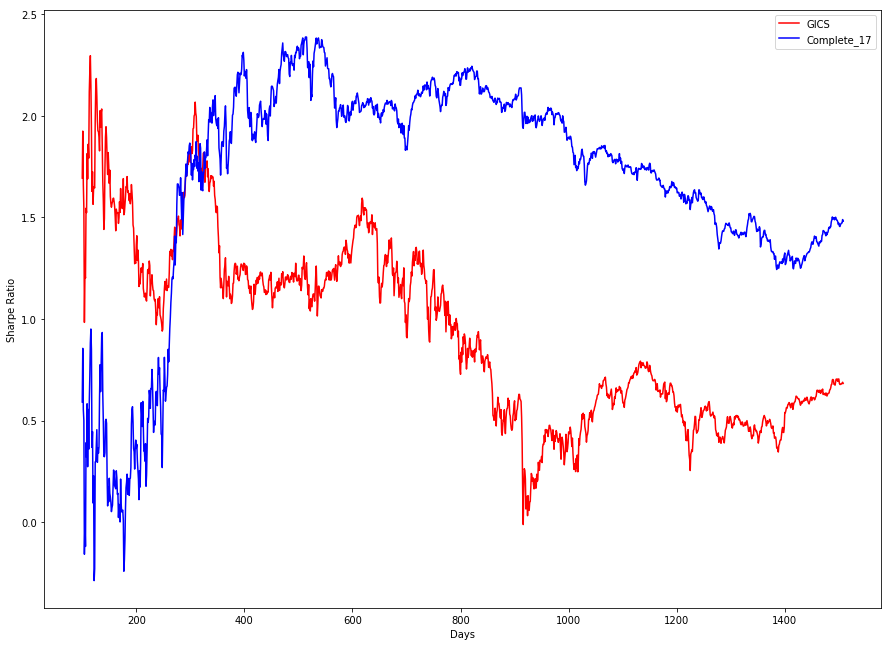} \\
        \textit{(c) Absolute Portfolio Value} & \textit{(d) Risk-adjusted Return} \\
        & \\
        \hline
    \end{tabular}
    % }
    \captionof{figure}{A comparison of sector universe performance metrics, with the \textcolor{red}{benchmark universe in red}, and the \textcolor{blue}{risk-adjusted return optimal learned sector universe in blue}.}
    \label{fig:benchmark_comparison:performance_metrics}
\end{table}

\pagebreak

\subsection{Absolute Portfolio Value Comparison}

Panel (c) in Figure~\ref{fig:benchmark_comparison:performance_metrics} is a comparison of the absolute portfolio value of both the risk-adjusted return optimal learned sector universe, and the benchmark universe. As indicated by the graph, the learned sector universe provides a significantly higher value at the terminus of the backtest, beating out the benchmark by nearly \$15,000,000,000 on a starting capital base of \$10,000,000,000 each, which translates to an outperformance of nearly 150\%.

The progression of the portfolio over time for both the learned sectors universe and the benchmark universe indicate that the portfolio returns of the two sector universes are lightly correlated. This is to be expected, as the underlying base of investable assets is identical (by design) between the two sector universes. However, there does appear to be significantly less historical volatility in the returns of the learned sector universe portfolio compared to the benchmark portfolio.

This is particularly evident in the 750 - 1250 day interval in panel (c). This period shows that the portfolio value of the benchmark increased rapidly, at a significantly greater rate than its learned sector universe counterpart. However, at approximately the 1150 day mark, the benchmark suffers a extremely severe drop, losing nearly all of its gains of the preceding period.

Interestingly, the learned sector universe portfolio does not appear to fluctuate in value significantly (relative to the benchmark) during this period. This observation, coupled with the commensurate final rally in both sector universe portfolios near the end of the backtesting period suggests that the diversification profile of the learned sectors portfolio is \textit{significantly} superior to that of the benchmark, resulting in not only a higher terminal portfolio value, but also significantly less volatility in reaching that value.

\subsection{Risk-Adjusted Return Comparison}

Figure~\ref{fig:benchmark_comparison:performance_metrics} (d) is a comparison of the rolling risk-adjusted return (i.e. Sharpe Ratio) of the benchmark sector universe and the risk-adjusted return optimal learned sector universe. Given the results of the analysis of the absolute portfolio value comparison above, the outperformance of the learned sector universe relative to the benchmark universe is not a surprising result.

Continuing on the same line of analysis as the previous section, the Sharpe ratio of the benchmark sector universe performs extremely poorly during the interval of 750 - 1250 days discussed above. The negative effect of the increased volatility, despite a rally in the underlying portfolio is better reflected in the Sharpe ratio plot compared to the portfolio value plot of panel (c). In fact, the Sharpe ratio graph in panel (d) indicates that it was nearly more beneficial to own and hold the risk-free asset than the benchmark sector universe portfolio at approximately the 900 day mark, as the rolling Sharpe ratio of the portfolio briefly approaches 0. Despite rallying significantly during the interval, the Sharpe ratio of the benchmark sector universe never recovers, and does not approach the significantly higher value of the learned sector universe portfolio.

Additionally, despite rallying toward the end of the backtest, the Sharpe ratio plots indicate that the trend of the rolling risk-adjusted return for both sector universes was negative, with a much more smooth slope on the learned sector universe portfolio line. This indicates a lower \textit{vol of vol} for the learned sector universe compared to the benchmark universe, which is further indication that the learned sector algorithm provides significantly better diversification benefits compared to the benchmark sector universe.

\section{Qualitative Comparison}

In this section, we attempt to conduct a more qualitatively-driven comparison and contrast of the risk-adjusted return optimal learned sector universe against the benchmark sector universe. Given the drastic difference in performance between the benchmark sector universe portfolio and the optimal learned sector universe portfolio, we believe that there is significant insight to be had by analyzing the composition of each of the sectors in the universe.

Analyzing each of the learned sectors in turn (from Figure~\ref{fig:optimal_sector_universe:max_sharpe} and Figure~\ref{fig:benchmark_comparison:ls_optimal_assets}), it is clear that beyond the large sectors \textit{Alpha} and \textit{Golf}, a large majority of the remaining sectors are extremely small with respect to their numbers of component assets. Despite this however, the two large (major) sectors - as well as a selection of the smaller (i.e. minor) sectors - have an extremely high dispersion rate relative to the benchmark. That is, there doesn't seem to be a high level of congruence between the old and new sector assignments. This lack of agreement between the benchmark and learned sector universes is particularly apparent in the apparent lack of any direct transitional sector mappings in Figure~\ref{fig:benchmark_comparison:ls_optimal_assets}.

\pagebreak

Both major learned sectors \textit{Alpha} and \textit{Golf} comprise a large number of assets as their components. Particularly, it can be observed that learned sector \textit{Alpha} contains a majority of the benchmark \textit{Financials} sector, and the benchmark \textit{Utilities} sector. Given that these sector assignments are derived from fundamentals data, this is a particularly interesting result, as both \textit{Financials} and \textit{Utilities} have become extremely risk-averse businesses over the last decade; \textit{Financials} due to the Great Recession of 2008, and \textit{Utilities} due to extensive capital damage incurred by the increased severity and number of Natural Disasters. This grouping indicates that the capital structure of these businesses are also becoming increasingly similar.

\begin{figure}[!h]
    \centering
    \fbox{
    \includegraphics[width=.7\linewidth]{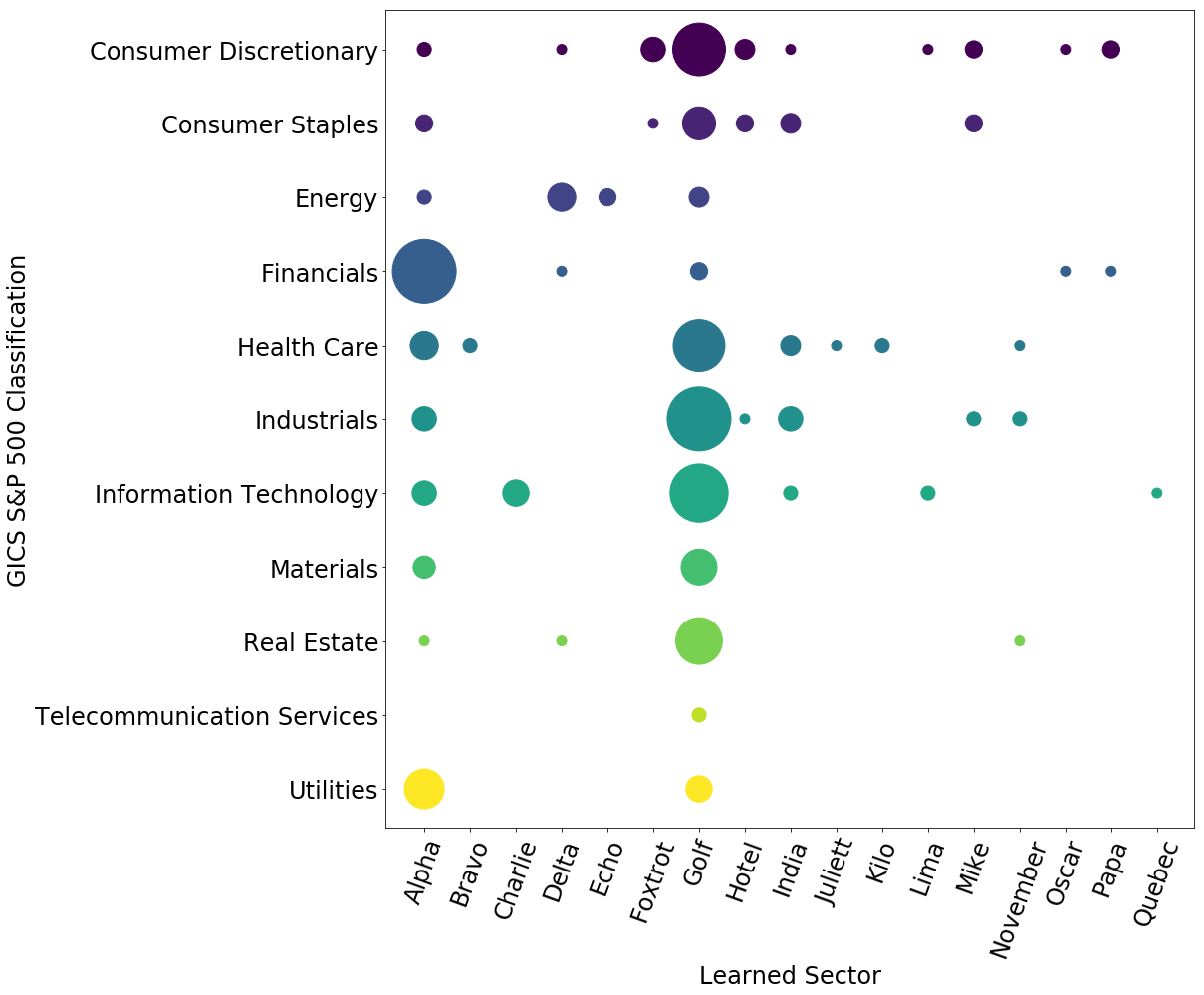}
    }
    \caption{Sector assignment transitions between the benchmark sector classification universe and the risk-adjusted return optimal learned sector universe.}
    \label{fig:benchmark_comparison:ls_optimal_assets}
\end{figure}

Considering learned sector \textit{Golf}, it seems to be a mini-index within the original sector universe. From a component count perspective, it ingests a large amount of the benchmark \textit{Consumer Discretionary} and \textit{Consumer Staples} industries, as well as large swaths of industries that form the backbone of the US Economy as a whole; namely, the \textit{Information Technology}, \textit{Industrials}, and \textit{Real Estate} sectors.

Appendix~\ref{appendix:portfolio_weights} contains stacked bar charts representing the level of investment in each of the sector SETFs for both the benchmark sector universe (see Figure~\ref{fig:appendix_weights:benchmark}), and the risk-adjusted return optimal learned sector universe (see Figure~\ref{fig:appendix_weights:sharpe_optimal}). Analyzing these graphs, in conjunction with the transition profile of sector assignments between the benchmark and optimal learned sector universe, it is clear that the learned sector \textit{Golf} was not a strong performer. The benefit of containing a large cross-section of companies from myriad traditional sectors, combined with their poor performance during the years of 2014 and 2015 appears to be a key factor in the outperformance of the benchmark sector universe.

% Conclusion

\chapter{Conclusion}
    
In this section, we will reiterate our main findings, and relate them back to our specific research goals and thesis statement, initially outlined in Section~\ref{research_goals}.

\vspace{-.5em}

% See: https://www.slideshare.net/linjaaho/how-to-make-boxed-text-with-latex
% See: ftp://ftp.dante.de/tex-archive/graphics/bclogo/doc/bclogo-doc.pdf [Its in French lmfao]
\begin{center}
    \begin{minipage}{0.7\textwidth}
        \begin{bclogo}[couleur=blue!30, arrondi=0.1, logo=\bcloupe, ombre=false]{\;Thesis Statement}
            Utilize relationships in the idiosyncratic characteristics of corporations to inform a fundamentals-driven, non-subjective sector classification framework.
        \end{bclogo}
    \end{minipage}
\end{center}

\vspace{-1.5em}

\section{Research Goal 1}

\begin{table}[h!]
    \centering
    \begin{tabular}{| c | c |}
        \hline
        &  \\
        RG-1 & Utilize data-driven algorithms to derive a truly objective classification heuristic. \\
        & \\
        \hline
    \end{tabular}
\end{table}

\begin{itemize}
    \item In the first portion of the report, we begin to address RG-1 goal by outlining our target data sources (see Section~\ref{model_data}), the benchmark we plan to use for comparison, and how our specifically selected fields from our data sources relate to, and reinforce our research objective.
    \item Following this, we recognized that in order to maintain the level of objectivity enforced by RG-1, we would have to use an unsupervised learning method to determine our candidate learned sector universes. To this end, we conducted a survey of potential methodologies and identified Hierarchical Clustering as our target methodology in Section~\ref{learning_methods_survey}.
    \item In Section~\ref{hierarchical_clustering_model}, we parameterized our HCA heuristic, and identified our search space consisting of 60 candidate learned sector universes. We then computed a set of candidate learned sector universes, fully addressing RG-1.
\end{itemize}

\section{Research Goal 2}

\begin{table}[h!]
    \centering
    \begin{tabular}{| c | c |}
        \hline
        &  \\
        RG-2 & Rank candidate sector universes against each other using entirely objective criteria. \\
        & \\
        \hline
    \end{tabular}
\end{table}

\begin{itemize}
    \item The second portion of the report was dedicated to addressing RG-2. This process began in Section~\ref{candidate_universe_ranking}, where we outlined the scope of RG-2. Following this, we introduced reIndexer, the backtest-driven sector universe evaluation research tool that powered the validation portion of our project (see Section~\ref{candidate_universe_ranking:reindexer}).
    \item Next, we utilized reIndexer to rank our candidate learned sector universes, and computed a set of performance metrics (see Section~\ref{candidate_universe_ranking:eval_metrics}) for each of our 60 candidate learned sector universes using reIndexer.
    \item Finally in Section~\ref{optimal_sector_universes}, we utilized these performance metrics to identify the risk-adjusted return optimal learned sector universe (Complete Linkage; 17 Sectors), and therefore completing RG-2.
\end{itemize}

\section{Research Goal 3}

\begin{table}[h!]
    \centering
    \begin{tabular}{| c | c |}
        \hline
        &  \\
        RG-3 & Evaluate our risk-adjusted return optimal sector universe against the benchmark. \\
        & \\
        \hline
    \end{tabular}
\end{table}

\begin{itemize}
    \item The final research goal of this report was addressed in Section~\ref{benchmark_comparison}.
    \item We compared the risk-adjusted return optimal learned sector universe to the benchmark (i.e. \textit{GICS S\&P 500 Classification} sector universe utilizing reIndexer, and the same performance metrics used to rank the candidate learned sector universes.
    \item Our comparison showed that the benchmark sector universe provided a lower level of both SETF restructuring and portfolio rebalancing turnover compared to the risk-adjusted return optimal learned sector universe. However, the learned sector universe significantly outperformed the benchmark universe with respect to total portfolio return and rolling risk-adjusted return.
    \item Following this, we conducted a thorough quantitative and qualitative analysis of the reIndexer output for both the risk-adjusted return optimal learned sector universe, and the benchmark sector universe.
    \item We conclude that our risk-adjusted optimal learned sector universe does indeed provide a superior diversification profile compared to the benchmark universe, thus fully realizing RG-3.
\end{itemize}

\vspace*{\fill}

\begin{center}
    \bfseries Having addressed our specific research goals, RG-1, RG-2, and RG-3, we assert and affirm that we fully explored the scope of the thesis statement of our FE 800 Project in Spring Semester 2019.
\end{center}

\vspace*{\fill}

% Future work

\chapter{Future Work}
    
In this section, we very briefly outline potential avenues for future research, building on the lessons learned during the course of this project.

\section{HCA Model Tuning}

Given the abundant pooling behavior (i.e. single large sector, and many single-asset sectors) of some of the hierarchical clustering models, it would be an extremely beneficial improvement to investigate methodologies to smooth the distribution of assets in the learned sectors.

\section{Varied ETF Construction Heuristics}

Currently, reIndexer creates and maintains price-weighted synthetic ETFs. However, a majority of market indexes today are market capitalization weighted, rather than price-weighted. A key improvement to reIndexer would be the implementation of market capitalization weighted SETFs, in addition to the current price-weighted SETF implementation.

\section{Temporal Variation of Sector Assignments}

As discussed in the report, we were unable to acquire historical sector assignment data for our benchmark sector universe, the \textit{GICS S\&P 500 Classification}. Due to this, we limited the scope of our sector ranking and benchmark evaluation to only use the latest fundamentals data we had available; 2017.

Given longitudinal sector assignment data for a collection of assets, reIndexer can be extended to be compatible with temporally varying sectors, increasing the overall accuracy of the evaluation system. This system would enable us to more accurately track metrics such as the SETF restructuring turnover over time, while also providing a more accurate assessment of the holding cost of SETFs to retail investors (i.e. portfolio rebalancing turnover).

\section{Existing Sectorization Scheme Ranking}

In addition to being an excellent tool for comparing hypothetical sector universes, reIndexer may also be used to compare existing sector classification schemes. That is, it may hypothetically be used to compare the performance of the GICS classification scheme against the FTSE classification scheme.

Similar to the analysis performed with the hypothetical sector universes, a \textit{diversification ranking} of sorts of existing sector universes may be developed.

% Appendices
\newpage

\appendix
\renewcommand\chaptername{Appendix}  % This is to change 'chapter' to 'appendix' in the header

% Uncomment these lines (up to \chapter) to suppress appendix name in the header
% Uncommenting leaves just chapter name and number, eg: "Appendix A"

% \fancypagestyle{plain}{
%     \fancyhead[LE,RO]{\chaptername \ \thechapter}
% }

% \fancypagestyle{appendix}{
%     \fancyhead[LE,RO]{\chaptername \ \thechapter}
% }

% \pagestyle{appendix}

\chapter{Backtest Visualization} \label{appendix:backtest_visualization}

\section*{SETF Restructuring Turnover}

See Figure~\ref{fig:appendix_backtest:setf_restructuring_turnover} on Page~\pageref{fig:appendix_backtest:setf_restructuring_turnover}.

\section*{Portfolio Rebalancing Turnover}

See Figure~\ref{fig:appendix_backtest:portfolio_rebal_turnover} on Page~\pageref{fig:appendix_backtest:portfolio_rebal_turnover}.

\section*{Portfolio Return}

See Figure~\ref{fig:appendix_backtest:portfolio_return} on Page~\pageref{fig:appendix_backtest:portfolio_return}.

\section*{Sharpe Ratio}

See Figure~\ref{fig:appendix_backtest:sharpe_ratio} on Page~\pageref{fig:appendix_backtest:sharpe_ratio}.

\newpage
\begin{sidewaysfigure}
    \centering
    \includegraphics[width=\linewidth]{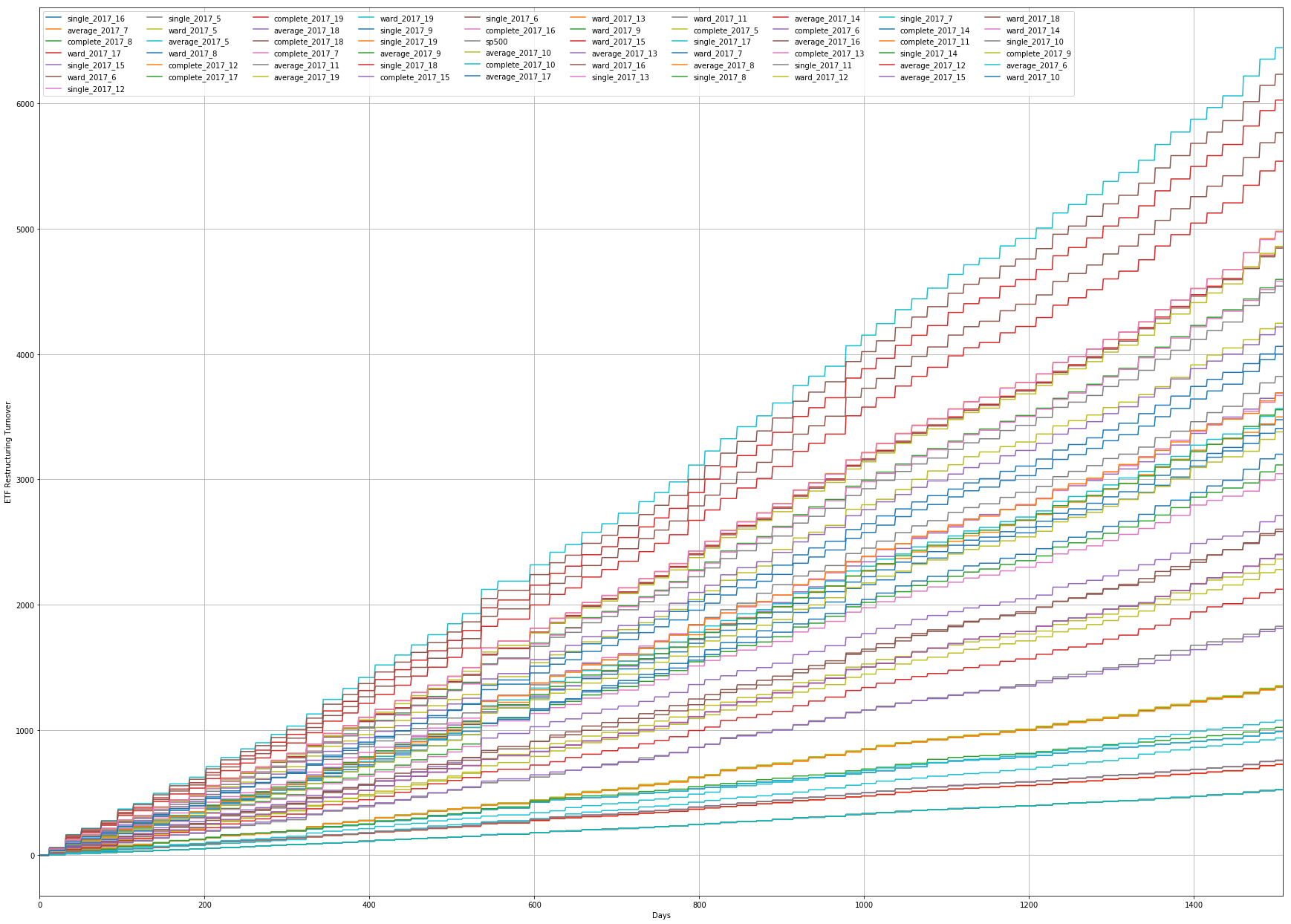}
    \caption{Cumulative SETF restructuring turnover for 60 candidate learned sector universes.}
    \label{fig:appendix_backtest:setf_restructuring_turnover}
    \addcontentsline{toc}{section}{SETF Restructuring Turnover}
\end{sidewaysfigure}

\newpage
\begin{sidewaysfigure}
    \centering
    \includegraphics[width=\linewidth]{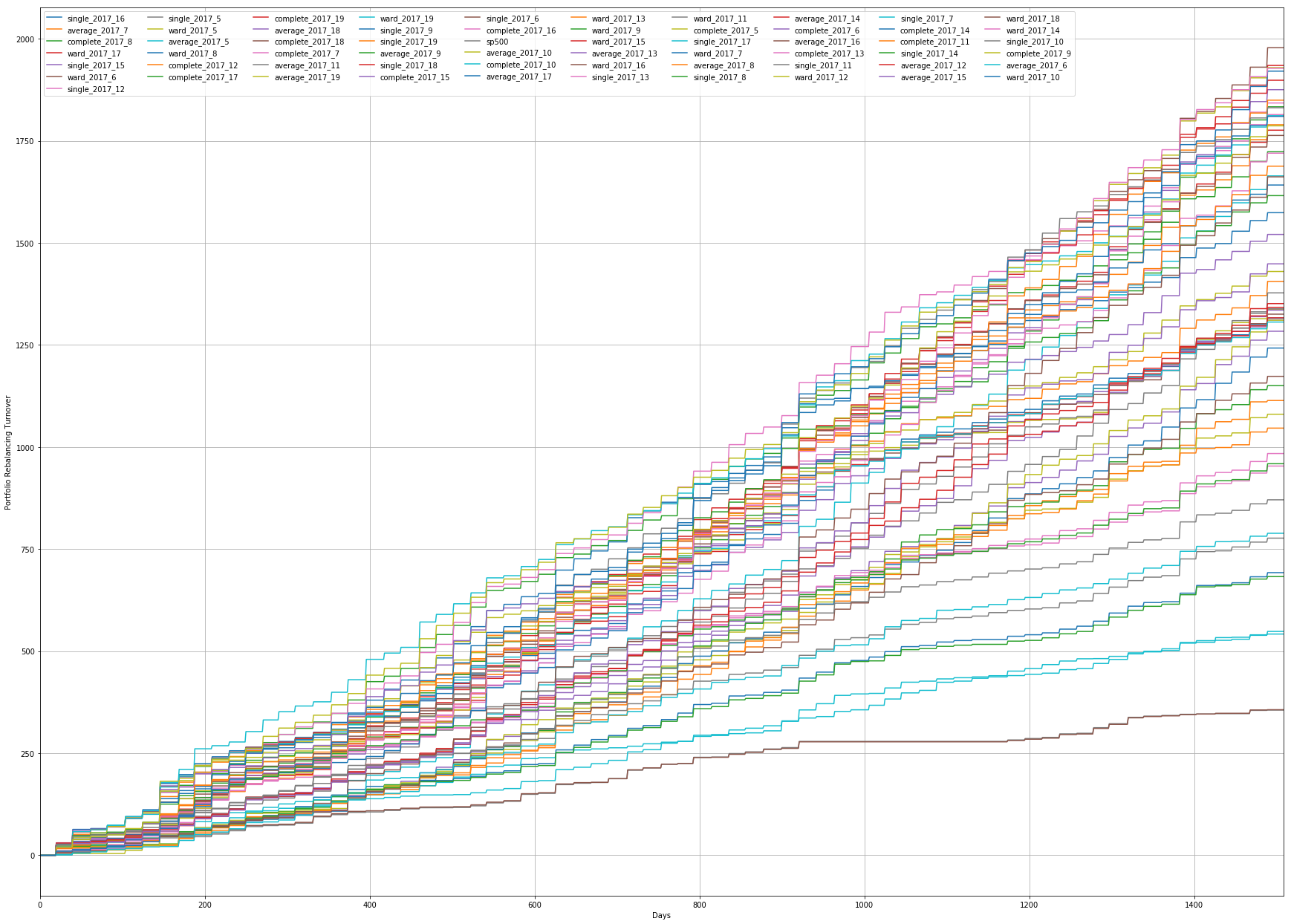}
    \caption{Cumulative portfolio rebalancing turnover for 60 candidate learned sector universes.}
    \label{fig:appendix_backtest:portfolio_rebal_turnover}
    \addcontentsline{toc}{section}{Portfolio Rebalancing Turnover}
\end{sidewaysfigure}

\newpage
\begin{sidewaysfigure}
    \centering
    \includegraphics[width=\linewidth]{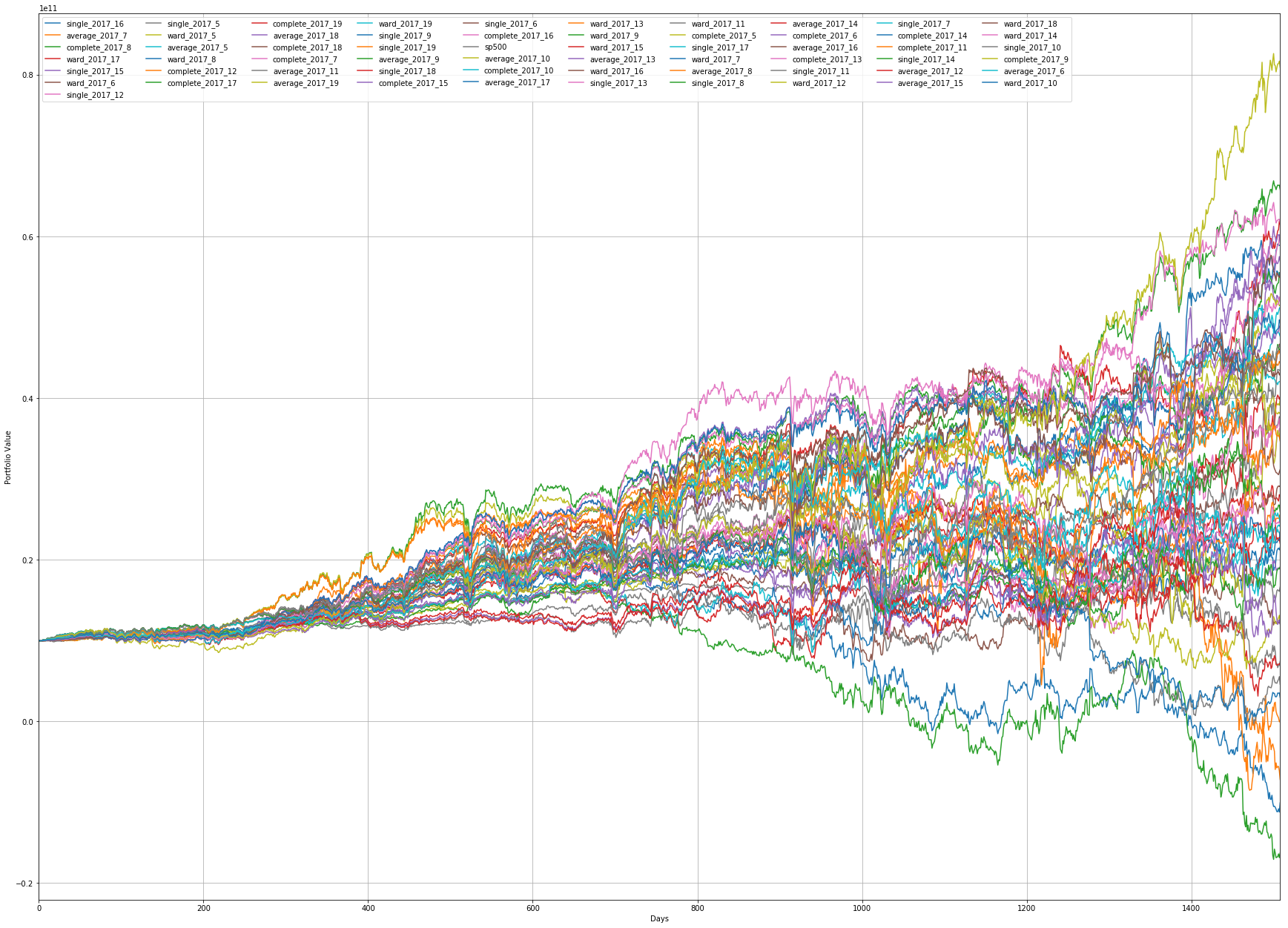}
    \caption{Portfolio return for 60 candidate learned sector universes.}
    \label{fig:appendix_backtest:portfolio_return}
    \addcontentsline{toc}{section}{Portfolio Return}
\end{sidewaysfigure}

\newpage
\begin{sidewaysfigure}
    \centering
    \includegraphics[width=\linewidth]{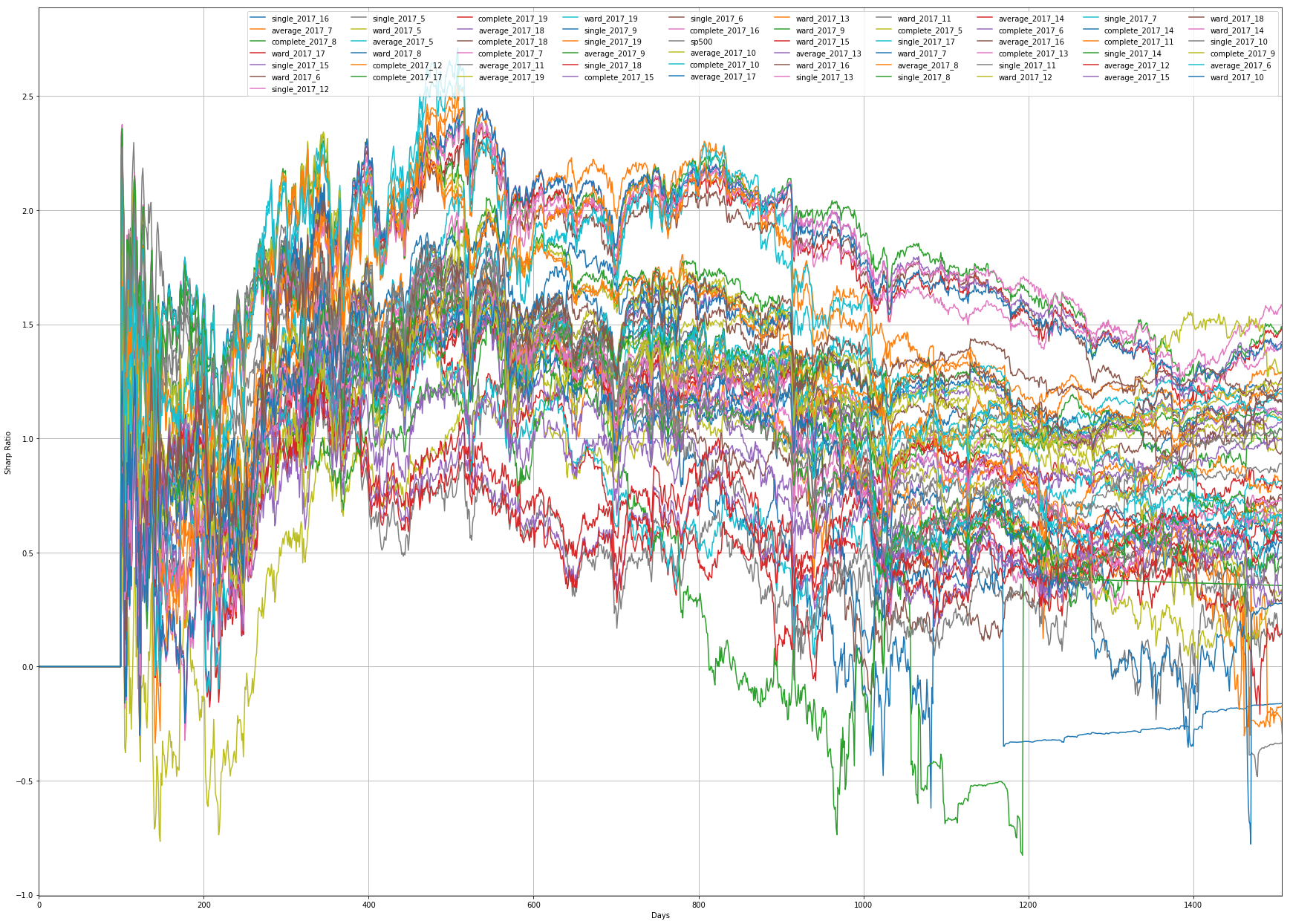}
    \caption{Rolling Sharpe Ratio for 60 candidate learned sector universes.}
    \label{fig:appendix_backtest:sharpe_ratio}
    \addcontentsline{toc}{section}{Sharpe Ratio}
\end{sidewaysfigure}

\chapter{Optimal Learned Sector Universe} \label{appendix:optimal_ls}

\section{Optimal Learned Sector Asset Distribution}

\begin{figure}[h]
    \centering
    \fbox{
    \includegraphics[width=.9\linewidth]{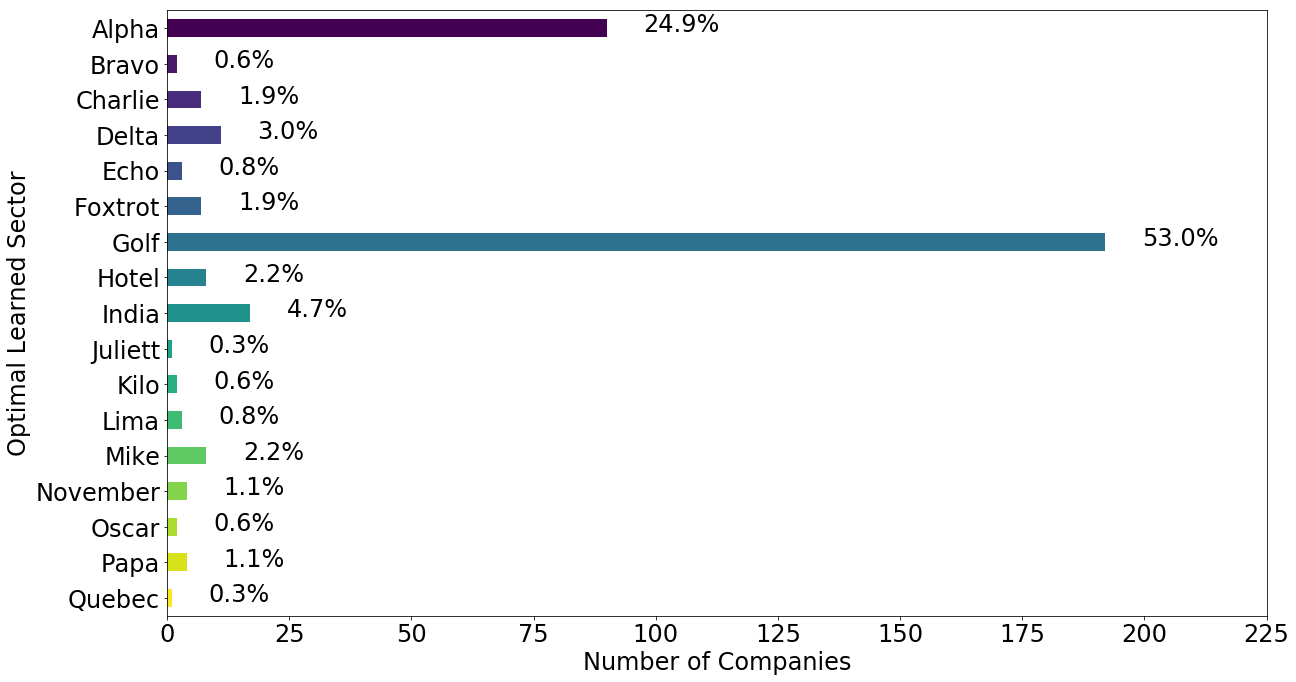}
    }
    \caption{Distribution of companies ($n = 362$) across sectors in the risk-adjusted return optimal learned sector universe.}
    \label{fig:appendix:optimal_ls:ls_sector_distribution}
\end{figure}

% Took forever to set up this table
% See: http://www.ctex.org/documents/packages/table/supertabular.pdf
% https://tex.stackexchange.com/questions/440078/how-can-i-generate-a-big-tabular-out-of-a-csv-file
% https://tex.stackexchange.com/questions/88387/disable-two-column-mode-for-separate-part
% https://tex.stackexchange.com/questions/269428/too-large-bottom-margin-with-xtab-or-supertabular
% https://latex.org/forum/viewtopic.php?t=5825
%
% Basically, supertabular over-estimates the page margins and causes a huge bottom margin to appear. To
% fix this the "\shrinkheight" command is run at each pagebreak (which is why its inserted into \tablehead)
% Had to seat first page height manually because this package is a mess
% Remember to add "showframe" option to geometry package when investigating this; figured out that it was
% messing up because the initial override I was setting was exceeding page margins

\twocolumn[\section{Optimal Learned Sector Universe Dataset}]

\begin{center}
    \tablefirsthead{%
        \shrinkheight{-0.5in}
        \hline
        \textbf{Ticker} & \textbf{GICS Sector} & \textbf{Optimal LS} \tabularnewline
        % \midrule
        \hline
    }
    \tablehead{%
        \shrinkheight{-1in}
        % \toprule
        \hline
        \textbf{Ticker} & \textbf{GICS Sector} & \textbf{Optimal LS} \tabularnewline
        % \midrule
        \hline
    }
    \tabletail{\hline}
    \tablelasttail{\hline}
    \begin{supertabular}{|c|c|c|}
        \begin{filecontents*}{report_data/complete_17_table.csv}
symbol,old_sector,new_sector
A,Health Care,Golf
AAP,Consumer Discretionary,Golf
AAPL,Information Technology,Golf
ABC,Health Care,Kilo
ABT,Health Care,Alpha
ACN,Information Technology,India
ADBE,Information Technology,Golf
ADI,Information Technology,Alpha
ADM,Consumer Staples,India
ADP,Information Technology,Alpha
ADS,Information Technology,Alpha
AEE,Utilities,Alpha
AEP,Utilities,Alpha
AES,Utilities,Golf
AGN,Health Care,Alpha
AIG,Financials,Alpha
AJG,Financials,Alpha
AKAM,Information Technology,Golf
ALGN,Health Care,Golf
ALK,Industrials,Golf
ALL,Financials,Alpha
ALXN,Health Care,Alpha
AMAT,Information Technology,Golf
AMD,Information Technology,Lima
AME,Industrials,Golf
AMG,Financials,Delta
AMGN,Health Care,Golf
AMZN,Consumer Discretionary,Lima
ANTM,Health Care,India
AON,Financials,Alpha
AOS,Industrials,Golf
APC,Energy,Delta
APD,Materials,Golf
APH,Information Technology,Golf
APTV,Consumer Discretionary,Golf
ARE,Real Estate,Golf
ATVI,Information Technology,Golf
AVB,Real Estate,Golf
AVGO,Information Technology,Golf
AVY,Materials,Golf
AWK,Utilities,Golf
AXP,Financials,Alpha
AYI,Industrials,Golf
AZO,Consumer Discretionary,Mike
BA,Industrials,Golf
BAC,Financials,Alpha
BAX,Health Care,Golf
BBT,Financials,Alpha
BBY,Consumer Discretionary,Hotel
BDX,Health Care,Alpha
BEN,Financials,Golf
BIIB,Health Care,Golf
BK,Financials,Alpha
BKNG,Consumer Discretionary,Foxtrot
BLL,Materials,Golf
BRK.B,Financials,Alpha
BSX,Health Care,Golf
BWA,Consumer Discretionary,Golf
BXP,Real Estate,Golf
C,Financials,Alpha
CAH,Health Care,Kilo
CAT,Industrials,Golf
CB,Financials,Alpha
CBOE,Financials,Alpha
CCI,Real Estate,Golf
CCL,Consumer Discretionary,Golf
CDNS,Information Technology,Lima
CERN,Health Care,Golf
CF,Materials,Golf
CHD,Consumer Staples,Golf
CHTR,Consumer Discretionary,Golf
CI,Health Care,Alpha
CINF,Financials,Alpha
CL,Consumer Staples,Mike
CLX,Consumer Staples,Mike
CMA,Financials,Alpha
CMCSA,Consumer Discretionary,Golf
CME,Financials,Alpha
CMG,Consumer Discretionary,India
CMI,Industrials,Golf
CMS,Utilities,Golf
CNC,Health Care,India
CNP,Utilities,Golf
COF,Financials,Alpha
COL,Industrials,Alpha
COO,Health Care,Golf
COP,Energy,Delta
COST,Consumer Staples,Hotel
CPB,Consumer Staples,Golf
CSCO,Information Technology,Golf
CSX,Industrials,Golf
CTSH,Information Technology,Golf
CTXS,Information Technology,Golf
CVS,Consumer Staples,India
CVX,Energy,Delta
CXO,Energy,Delta
DAL,Industrials,Golf
DE,Industrials,Alpha
DFS,Financials,Alpha
DGX,Health Care,Golf
DHI,Consumer Discretionary,Papa
DHR,Health Care,Alpha
DIS,Consumer Discretionary,Golf
DISCA,Consumer Discretionary,Golf
DISH,Consumer Discretionary,Delta
DLR,Real Estate,Golf
DOV,Industrials,Golf
DRE,Real Estate,Golf
DTE,Utilities,Alpha
DUK,Utilities,Alpha
DVA,Health Care,Golf
DWDP,Materials,Alpha
EBAY,Information Technology,Golf
ECL,Materials,Golf
ED,Utilities,Alpha
EFX,Industrials,Golf
EIX,Utilities,Alpha
EL,Consumer Staples,Foxtrot
EMR,Industrials,Golf
EOG,Energy,Echo
EQIX,Real Estate,Golf
EQR,Real Estate,Golf
ES,Utilities,Alpha
ESS,Real Estate,Golf
ETFC,Financials,Alpha
ETN,Industrials,Golf
ETR,Utilities,Alpha
EW,Health Care,Golf
EXC,Utilities,Alpha
EXPD,Industrials,India
EXPE,Consumer Discretionary,Foxtrot
F,Consumer Discretionary,Golf
FAST,Industrials,Mike
FB,Information Technology,Charlie
FBHS,Industrials,Golf
FCX,Materials,Golf
FE,Utilities,Golf
FFIV,Information Technology,Charlie
FIS,Information Technology,Golf
FISV,Information Technology,Golf
FITB,Financials,Alpha
FLIR,Information Technology,Golf
FLR,Industrials,India
FLS,Industrials,Golf
FMC,Materials,Alpha
FOXA,Consumer Discretionary,Golf
FRT,Real Estate,Golf
GD,Industrials,Golf
GILD,Health Care,Golf
GLW,Information Technology,Golf
GM,Consumer Discretionary,Golf
GPC,Consumer Discretionary,Golf
GPN,Information Technology,Alpha
GRMN,Consumer Discretionary,Golf
GT,Consumer Discretionary,Golf
GWW,Industrials,Hotel
HAL,Energy,Golf
HAS,Consumer Discretionary,Foxtrot
HBAN,Financials,Alpha
HBI,Consumer Discretionary,Golf
HCA,Health Care,November
HCP,Real Estate,Golf
HD,Consumer Discretionary,Mike
HES,Energy,Echo
HIG,Financials,Alpha
HII,Industrials,Golf
HOLX,Health Care,Golf
HON,Industrials,Golf
HP,Energy,Echo
HPQ,Information Technology,India
HRL,Consumer Staples,Golf
HRS,Information Technology,Golf
HSIC,Health Care,Golf
HST,Real Estate,Golf
HSY,Consumer Staples,Mike
HUM,Health Care,India
IBM,Information Technology,Golf
ICE,Financials,Alpha
IDXX,Health Care,Golf
ILMN,Health Care,Golf
INCY,Health Care,Bravo
INFO,Industrials,Alpha
INTC,Information Technology,Golf
INTU,Information Technology,Charlie
IP,Materials,Golf
IPG,Consumer Discretionary,Alpha
IPGP,Information Technology,Golf
IR,Industrials,Golf
IRM,Real Estate,Golf
ISRG,Health Care,Golf
IT,Information Technology,Alpha
ITW,Industrials,Golf
JBHT,Industrials,India
JCI,Industrials,Golf
JEC,Industrials,India
JNJ,Health Care,Golf
JNPR,Information Technology,Golf
JPM,Financials,Alpha
JWN,Consumer Discretionary,Hotel
K,Consumer Staples,Golf
KEY,Financials,Alpha
KIM,Real Estate,Golf
KLAC,Information Technology,Golf
KMB,Consumer Staples,Golf
KMI,Energy,Golf
KO,Consumer Staples,Golf
KSU,Industrials,Golf
LEG,Consumer Discretionary,Golf
LEN,Consumer Discretionary,Papa
LH,Health Care,Alpha
LKQ,Consumer Discretionary,Golf
LLL,Industrials,Golf
LLY,Health Care,Golf
LMT,Industrials,Golf
LNT,Utilities,Alpha
LOW,Consumer Discretionary,Hotel
LRCX,Information Technology,Golf
LUV,Industrials,Golf
LYB,Materials,Golf
M,Consumer Discretionary,Golf
MA,Information Technology,Charlie
MAC,Real Estate,Golf
MAS,Industrials,Golf
MAT,Consumer Discretionary,Foxtrot
MCD,Consumer Discretionary,Golf
MDLZ,Consumer Staples,Alpha
MET,Financials,Alpha
MGM,Consumer Discretionary,Golf
MHK,Consumer Discretionary,Golf
MKC,Consumer Staples,Alpha
MMC,Financials,Golf
MMM,Industrials,Golf
MNST,Consumer Staples,Golf
MO,Consumer Staples,Golf
MOS,Materials,Alpha
MRO,Energy,Delta
MSFT,Information Technology,Golf
MSI,Information Technology,Golf
MTB,Financials,Alpha
MTD,Health Care,Golf
MU,Information Technology,Golf
NAVI,Financials,Alpha
NBL,Energy,Delta
NCLH,Consumer Discretionary,Golf
NDAQ,Financials,Alpha
NEE,Utilities,Golf
NEM,Materials,Golf
NFLX,Information Technology,Alpha
NI,Utilities,Alpha
NKTR,Health Care,Juliett
NLSN,Industrials,Golf
NOC,Industrials,Alpha
NOV,Energy,Alpha
NSC,Industrials,Golf
NTRS,Financials,Alpha
NUE,Materials,Golf
NVDA,Information Technology,Charlie
NWL,Consumer Discretionary,Golf
O,Real Estate,Golf
OMC,Consumer Discretionary,Alpha
ORLY,Consumer Discretionary,Mike
OXY,Energy,Delta
PBCT,Financials,Alpha
PCAR,Industrials,Alpha
PEP,Consumer Staples,Golf
PG,Consumer Staples,Golf
PH,Industrials,Golf
PHM,Consumer Discretionary,Papa
PKI,Health Care,Alpha
PLD,Real Estate,Delta
PM,Consumer Staples,Golf
PNR,Industrials,Golf
PNW,Utilities,Alpha
PPG,Materials,Golf
PPL,Utilities,Golf
PRGO,Health Care,Golf
PSX,Energy,Delta
PWR,Industrials,India
QCOM,Information Technology,Golf
RCL,Consumer Discretionary,Golf
RE,Financials,Alpha
REG,Real Estate,Golf
REGN,Health Care,Golf
RF,Financials,Alpha
RHI,Industrials,Mike
RJF,Financials,Alpha
RMD,Health Care,Golf
ROK,Industrials,Golf
ROP,Industrials,Alpha
RSG,Industrials,Golf
RTN,Industrials,Golf
SBAC,Real Estate,November
SBUX,Consumer Discretionary,Oscar
SCHW,Financials,Papa
SEE,Materials,Golf
SHW,Materials,Golf
SIVB,Financials,Alpha
SLB,Energy,Alpha
SLG,Real Estate,Golf
SNA,Consumer Discretionary,Golf
SNPS,Information Technology,Golf
SO,Utilities,Alpha
SPG,Real Estate,Golf
SPGI,Financials,Golf
SRE,Utilities,Alpha
STI,Financials,Alpha
STT,Financials,Alpha
STX,Information Technology,Golf
SWK,Consumer Discretionary,Golf
SWKS,Information Technology,Charlie
SYK,Health Care,Golf
SYY,Consumer Staples,Hotel
T,Telecommunication Services,Golf
TAP,Consumer Staples,Alpha
TDG,Industrials,November
TEL,Information Technology,Golf
TPR,Consumer Discretionary,Golf
TRIP,Consumer Discretionary,Foxtrot
TROW,Financials,Oscar
TRV,Financials,Alpha
TSCO,Consumer Discretionary,Hotel
TSN,Consumer Staples,India
TSS,Information Technology,Golf
TXN,Information Technology,Charlie
TXT,Industrials,Golf
UAA,Consumer Discretionary,Foxtrot
UAL,Industrials,Golf
UDR,Real Estate,Alpha
UHS,Health Care,Golf
UNH,Health Care,India
UNM,Financials,Alpha
UNP,Industrials,Golf
UPS,Industrials,India
URI,Industrials,November
USB,Financials,Alpha
UTX,Industrials,Golf
V,Information Technology,Golf
VAR,Health Care,Golf
VMC,Materials,Alpha
VNO,Real Estate,Golf
VRSK,Industrials,Golf
VRSN,Information Technology,Quebec
VRTX,Health Care,Bravo
VTR,Real Estate,Golf
VZ,Telecommunication Services,Golf
WBA,Consumer Staples,India
WDC,Information Technology,Golf
WEC,Utilities,Alpha
WFC,Financials,Alpha
WHR,Consumer Discretionary,Golf
WM,Industrials,Golf
WMB,Energy,Golf
WMT,Consumer Staples,Hotel
WRK,Materials,Alpha
WU,Information Technology,Golf
WY,Real Estate,Golf
XEC,Energy,Golf
XEL,Utilities,Alpha
XRX,Information Technology,Golf
XYL,Industrials,Golf
ZION,Financials,Alpha
ZTS,Health Care,Golf
\end{filecontents*}
\csvreader[
            late after line=\\
        ]{report_data/complete_17_table.csv}{1=\ticker, 2=\gics, 3=\ls}{\ticker & \gics & \ls}
    \end{supertabular}
\end{center}

\onecolumn

\chapter{Backtest Portfolio Weights} \label{appendix:portfolio_weights}

\section*{Optimal Learned Sector Universe Portfolio}

See Figure~\ref{fig:appendix_weights:sharpe_optimal} on Page~\pageref{fig:appendix_weights:sharpe_optimal}.

\section*{Benchmark Sector Universe Portfolio}

See Figure~\ref{fig:appendix_weights:benchmark} on Page~\pageref{fig:appendix_weights:benchmark}.

\newpage
\begin{sidewaysfigure}
    \centering
    \includegraphics[width=\linewidth]{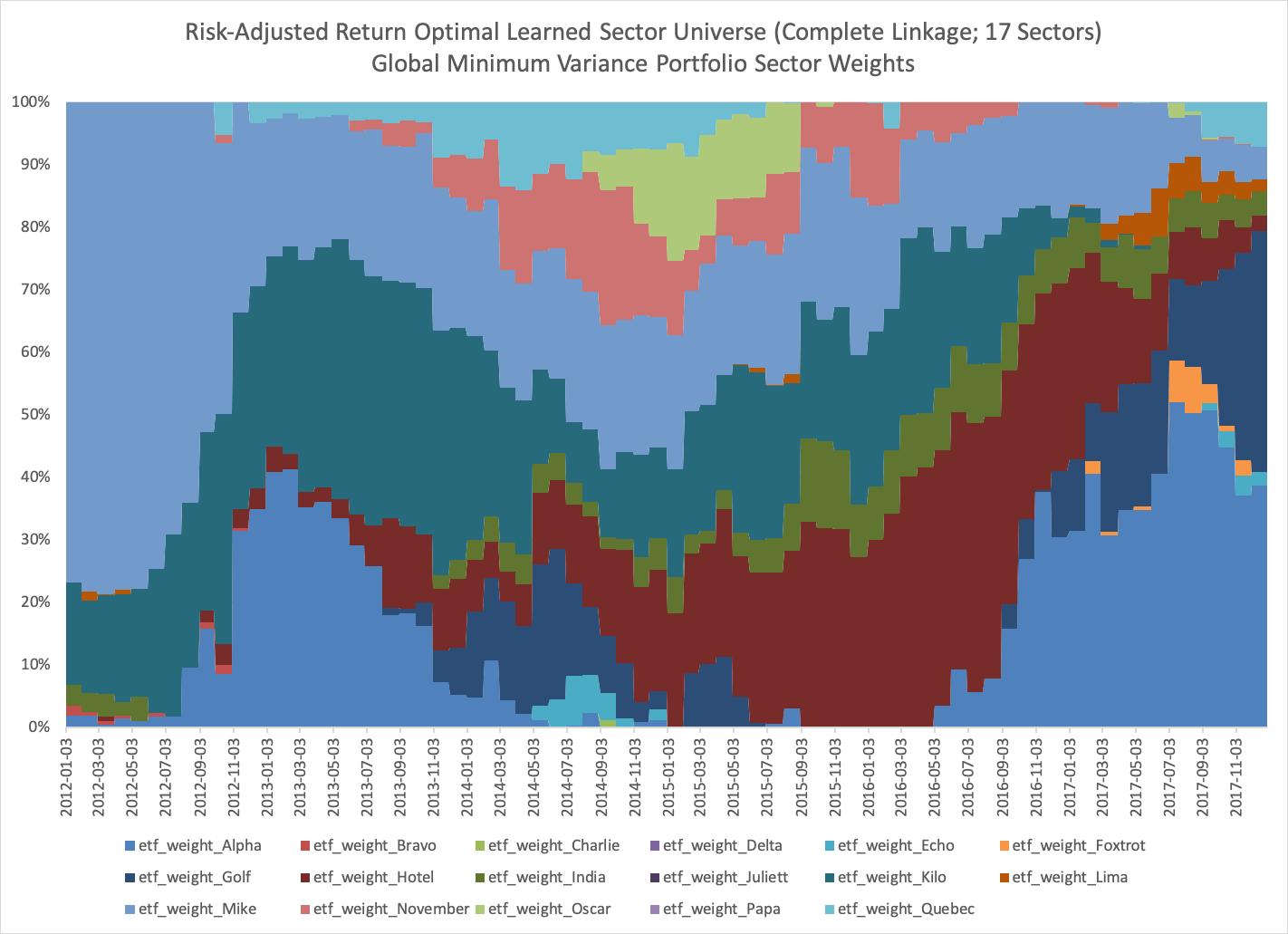}
    \caption{Risk-adjusted return optimal learned sector universe portfolio assignment weights.}
    \label{fig:appendix_weights:sharpe_optimal}
    \addcontentsline{toc}{section}{Optimal Learned Sector Universe Portfolio}
\end{sidewaysfigure}

\newpage
\begin{sidewaysfigure}
    \centering
    \includegraphics[width=\linewidth]{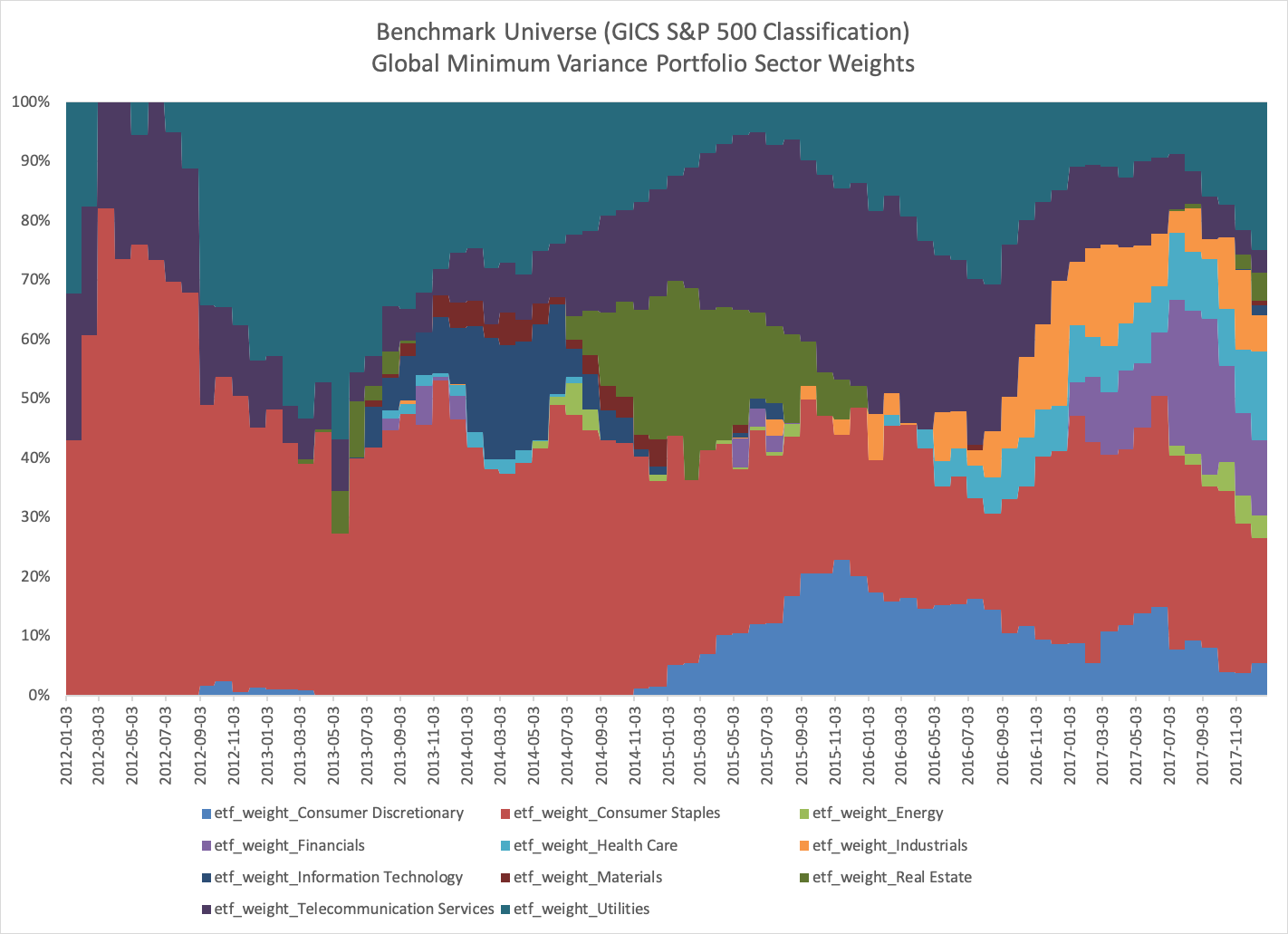}
    \caption{Benchmark sector universe (i.e. \textit{GICS S\&P 500 Classification}) portfolio assignment weights.}
    \label{fig:appendix_weights:benchmark}
    \addcontentsline{toc}{section}{Benchmark Sector Universe Portfolio}
\end{sidewaysfigure}

% References
\newpage
\nocite{Weerawarana2019FECode} % Report source code
\addcontentsline{toc}{chapter}{Bibliography}
\printbibliography

% Glossary
\glsaddall
\printglossary[style=altlist, nonumberlist]

\end{document}